\documentclass{article}

\usepackage[nonatbib, preprint]{neurips_2024}

\usepackage{amsmath,amsthm,amssymb,amsfonts, fancyhdr, color, comment, graphicx, environ}
\usepackage[square,numbers]{natbib}
\usepackage{hyperref}
\usepackage{xcolor}
\usepackage{mdframed}
\usepackage[shortlabels]{enumitem}
\usepackage{indentfirst}
\usepackage{hyperref}
\usepackage{rotating}
\usepackage{listings}
\usepackage{braket}
\usepackage{booktabs}
\usepackage{float}
\usepackage{tcolorbox}
\usepackage{algorithm}
\usepackage[algo2e]{algorithm2e} 
\usepackage{algpseudocode}
\usepackage{multirow}
\numberwithin{equation}{section}

\newtheorem{theorem}{Theorem}[section]
\newtheorem{lemma}[theorem]{Lemma}

\usepackage[utf8]{inputenc} 
\usepackage[T1]{fontenc}    
\usepackage{hyperref}       
\usepackage{url}            
\usepackage{booktabs}       
\usepackage{amsfonts}       
\usepackage{nicefrac}       
\usepackage{microtype}      
\usepackage{xcolor}         

\title{Statistical Arbitrage in Rank Space}

\author{
  Ying-Fei Li\\
  Department of Applied Physics\\
  Department of Statistics\\
  Stanford University\\
  Stanford, CA 94305 \\
  \texttt{yingfeil@stanford.edu} \\
  \And
  George C. Papanicolaou\\
  Department of Mathematics \\
  Stanford University\\
  Stanford, CA, 94305\\
  \texttt{papanicolaou@stanford.edu}
}

\begin{document}
\maketitle
\begin{abstract}
Equity market dynamics are conventionally investigated in name space, where stocks are indexed by company names. However, this perspective often suffers from high volatility and a low signal-to-noise ratio, which poses challenges for effective learning by deep neural networks (DNNs). In contrast, by indexing stocks by their ranks in capitalization, we gain a distinct and more structured view of market behavior in rank space. In this work, we demonstrate that DNNs achieve superior performance in statistical arbitrage when operating in rank space compared to name space. This performance gain is driven by more robust market representations and enhanced mean-reverting properties of residual returns in rank space, which facilitate more efficient learning. Our findings highlight the critical role of domain-informed data transformation in improving deep learning performance in noisy financial environments.
\end{abstract}

\section{Introduction}

In equity markets, stocks are conventionally labeled by fixed company identifiers (company names), a perspective that often suffers from high volatility and a low signal-to-noise ratio, hindering effective learning by deep neural networks (DNNs). 

As an alternative, we consider a rank-space representation, where stocks are indexed by their ranks in capitalization rather than fixed company identifiers. In this view, we focus on the stock at a given rank in capitalization while the corresponding company name may change. This formulation offers a more structured and stable view of market dynamics. We refer to a market labeled by fixed company identifiers (company names) as the \textit{market in name space}, and one labeled by ranks in capitalization as the \textit{market in rank space}.

A structured market dynamics is crucial for the performance of statistical arbitrage, a market-neutral strategy that seeks to exploit temporary mispricing among related assets\cite{avellaneda2010statistical, guijarro2021deep}. Specifically, it constructs portfolios that rely on the residual returns, the component of returns unexplained by market factors, and profits if these residuals exhibit mean-reverting behavior. Therefore, the presence of structured market dynamics, particularly mean reversion in residuals, is crucial to its performance.

In this paper, we show that operating statistical arbitrage in rank space by DNNs significantly outperforms the name-space counterpart. Our DNN-based rank-space portfolios achieve an average annual return 35.68\% and an average Sharpe ratio of 3.28 from 2007 to 2022, accounting for a 2-basis-point transaction cost. In contrast, applying the same DNNs in name space yields negligible returns over the same period.

Further comparison shows that DNN-based rank-space portfolios not only exploit the importance of mean-reversion during mean-variance optimization, but also implement more intelligent strategies than traditional parametric model by applying flexible leverage and minimizing carry risk. We attribute these improvements to two key advantages of rank space: (i) a more robust and stable representation of market structure, and (ii) enhanced mean-reverting behavior in residual returns. Together, these properties enable DNNs to extract actionable signals from noisy financial data.

Our findings highlight the importance of domain-informed data representations in financial machine learning. In particular, they show how appropriate input transformation can significantly improve the learning efficiency and performance of deep models in complex, noisy environments like equity markets\cite{bouchaud2009financial, pafka2003noisy}.

The remainder of the paper is organized as follows. Section 2 reviews related work. Section 3 formulates the framework for statistical arbitrage in both name space and rank space. Section 4 presents our empirical results using U.S. equity date. Section 5 concludes with a discussion of implications and future directions.

\section{Related work}
Our results add to the burgeoning literature on machine learning applications in finance. Gu \textit{et al.} systematically compare various machine learning methods for predicting stock returns\cite{gu2020empirical}. Acero \textit{et al.} utilize deep reinforcement learning for portfolio managements\cite{acero2024deep}. Araci explores the correlation between financial news and stock returns by BERT language model\cite{araci2019finbert}. Horvath \textit{et al.} apply deep neural networks to calibrate volatility surface in fixed-income markets\cite{horvath2021deep}.

Our research advances the understanding of statistical arbitrage. While this problem can be framed as a stochastic control problem solvable via the Hamilton-Jacobi-Bellman (HJB) equation\cite{cartea2015algorithmic, leung2015optimal, tourin2013dynamic, yong2012stochastic}, practical applications are often limited by calibration challenges and the absence of selection mechanism in diverse markets. To address these issues, Avellaneda and Lee\cite{avellaneda2010statistical} propose a pragmatic trading strategy based on Ornstein–Uhlenbeck (OU) processes, later refined by Yeo and Papanicolaou\cite{yeo2017risk}. More recently, Mulvey \textit{et al.}\cite{mulvey2020optimizing} combine the HJB framework with feed-forward networks, while Kim \textit{et al.}\cite{kim2019optimizing} apply deep reinforcement learning to optimize pair trading strategies. Guijarro-Ordonez \textit{et al.}\cite{guijarro2021deep} provide a comprehensive survey of deep learning models for statistical arbitrage.

Our paper also enriches the emerging literature on rank-based market models. Fernholz \textit{et al.}, in their seminal work on stochastic portfolio theory, introduce the concept of functionally generated portfolios in both name and rank spaces\cite{fernholz2002stochastic}. Building on this foundation, Banner \textit{et al.}\cite{banner2005atlas} and Ichiba \textit{et al.}\cite{ichiba2011hybrid} study the dynamics of rank capitalizations and gap processes in hybrid Atlas models, starkly contrasting the dynamics in name space. Healy \textit{et al.}\cite{bhealy2025} demonstrate that rank space exhibits a more structured market in rank space, theoretically establishing a larger spectral gap in the correlation matrix compared to name space, supported by comprehensive empirical validation (Figure \ref{fig: market_structure_name_vs_rank}). These theoretical insights form the basis for our proposed statistical arbitrage strategy in rank space.

\section{Formulation}
This section formulates statistical arbitrage in name and rank space. A schematic overview is shown in \autoref{fig: schematic_formulation_framework}. Supplementary implementation details in Appendix~\ref{sec: implementation_details}, with detailed algorithmic pseudocode is provided in Appendix~\ref{section: algorithms}. The code is available at \url{https://github.com/Infi-Yingfei-Li/stats-arb-rank-space}. 

\begin{figure}
    \centering
    \includegraphics[scale=0.22]{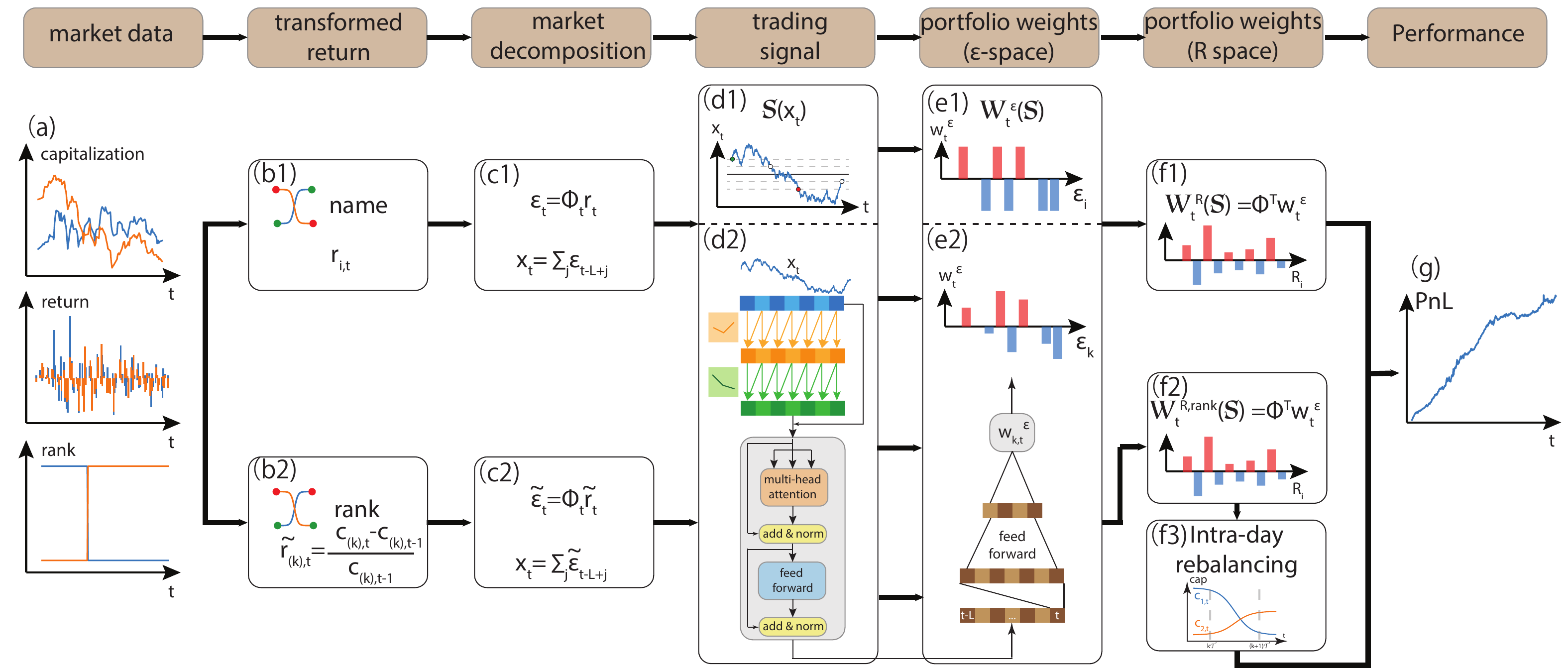}
    \caption{Schematic of the statistical arbitrage algorithm in name space and rank space.}
    \label{fig: schematic_formulation_framework}
\end{figure}

\subsection{Market decomposition}\label{section: market_decomposition}
\subsubsection{Name space}
In a market consisting of $N$ stocks, we denote the dividend-adjusted return\footnote{The daily return of a stock that accounts for both price changes and dividend payments.} on stock $i$ at trading day $t$ by $r_{i,t}$. We adopt a standard factor model for stock returns, 
\begin{equation}\label{eq: market_decomposition_name_main}
    r_t - r_f = \beta_t F_t + \epsilon_t, \quad \quad t=1,2, ..., T.
\end{equation}
Here, $r_t = \{r_{i,t}\}_{i=1}^N \in \mathbb{R}^{N}$ are the dividend-adjusted daily return, $r_f \in \mathbb{R}$ is the risk-free rate\footnote{The investment return with zero-risk financial loss, chosen as the rate of 1-month U.S. treasury bill here.}, $F_t\in \mathbb{R}^{K\times 1}$ are the underlying factors, $\beta_t\in \mathbb{R}^{N\times K}$ are the corresponding loadings on $K$ factors, and $\epsilon_t\in\mathbb{R}^{N}$ are the residual returns. 

Without loss of generality, these factors can be written as portfolios of stocks,
\begin{equation}\label{eq: market_decomposition_name_factor}
    F_t = \omega_t (r_t - r_f),
\end{equation}
where $\omega_t \in \mathbb{R}^{K\times N}$ specifies the factor portfolio weights. Combining \eqref{eq: market_decomposition_name_main} and \eqref{eq: market_decomposition_name_factor} yields
\begin{equation}
    r_t - r_f = \beta_t \omega_t (r_t - r_f) + \epsilon_t \Rightarrow \epsilon_t = (I-\beta_t\omega_t)(r_t-r_f) := \Phi_t (r_t-r_f)
\end{equation}
where $\Phi_t := (I - \beta_t \omega_t)$ defines a linear transformation from returns to residual returns. Each $\epsilon_{i,t}$ can thus be interpreted as the return of a tradable portfolio with weights given by the $i$-th row of $\Phi_t$. 

Consequently, we refer to the space spanned by $r_t$ as the \textit{name equity space}, and the space spanned by $\epsilon_t$ as the \textit{name residual space}. We denote the portfolio weights in name equity space as $w_t^{R, \text{name}}$ and portfolio weights in name residual space as $w_t^{\epsilon, \text{name}}$. These are related by 
\begin{equation}\label{eq: portfolio_weights_conversion_name_space}
    w_t^{R, \text{name}} = \Phi_t^Tw_t^{\epsilon, \text{name}}
\end{equation}
Importantly, given any $w_t^{\epsilon, \text{name}}$, the derived $w_t^{R, \text{name}}$ are market neural (proofs in the Appendix \ref{section: market_neutrality}).

\subsubsection{Rank space}
We begin by introducing the key variables that characterize the market dynamics in rank space: the daily return on rank $k$ at day $t$ in the continuous-time limit, defined as
\begin{equation}\label{eq: def_return_rank}
    \Tilde{r}_{(k),t} := \frac{c_{(k), t} - c_{(k), t-1}}{c_{(k), t-1}} = \frac{c_{\mathcal{I}_{(k),t}, t} - c_{\mathcal{I}_{(k),t-1}, t-1}}{c_{\mathcal{I}_{(k),t-1}, t-1}},
\end{equation}
where $c_{i,t}$ is the capitalization of stock $i$ at day $t$, $c_{(k),t}$ is the capitalization of the stock occupying the $k$-th rank in descending order at day $t$. $\mathcal{I}_{(k), t}$ represents the company occupying the rank $k$ at day $t$, and conversely, $\mathcal{R}_{i,t}$ gives the capitalization rank of stock $i$. Our definition of return in rank space is motivated by the log capitalization process in the hybrid-Atlas model in Appendix \ref{sec: hybrid-atlas model}\cite{banner2005atlas, ichiba2011hybrid}.

Importantly, $\Tilde{r}_t$ does not necessarily correspond to direct financial quantity, as $\mathcal{I}_{(k),t}$ and $\mathcal{I}_{(k), t-1}$ may differ -- meaning the stock occupying rank $k$ can change between days. To realize $\Tilde{r}_t$ in practice, we develop an intra-day rebalancing strategy, detailed in Appendix \ref{section: intraday_rebalancing}. 

Following the construction in name space, we assume a factor model for $\Tilde{r}_t$,
\begin{equation}\label{eq: market_decomposition_PCA_rank_main}
    \Tilde{r}_t - r_f = \Tilde{\beta_t} \Tilde{F_t} + \Tilde{\epsilon}_t,
\end{equation}
where $\Tilde{r}_t = \{r_{(k), t}\}_{k=1}^N \in \mathbb{R}^{N}$, $\Tilde{\beta_t} \in \mathbb{R}^{N\times K}$, $\Tilde{F_t} \in \mathbb{R}^{K\times 1}$, and $\Tilde{\epsilon}_t\in \mathbb{R}^{N}$. Analogously, we model the factors as portfolios of rank returns
\begin{equation}\label{eq: market_decomposition_PCA_rank}
    \Tilde{r}_t - r_f = \Tilde{\beta}_t\Tilde{\omega}_t (\Tilde{r}_t - r_f) + \Tilde{\epsilon}_t \Rightarrow \Tilde{\epsilon}_t = (I-\Tilde{\beta}_t\Tilde{\omega}_t)(\Tilde{r}_t - r_f) := \Tilde{\Phi}_t (\Tilde{r}_t - r_f),
\end{equation}
where $\Tilde{\Phi}_t:=(I-\Tilde{\beta}_t\Tilde{\omega}_t)$. 

We refer to the space spanned by $\Tilde{r}_t$ as \textit{rank equity space} and the space spanned by $\Tilde{\epsilon}_t$ as \textit{rank residual space}, mirroring our definition in the name space. Let $w_t^{R, \text{rank}}$ and $w_t^{\epsilon, \text{rank}}$ denote the portfolio weights in rank equity space and rank residual space, respectively, related by
\begin{equation}\label{eq: portfolio_weights_conversion_rank_space}
    w_t^{R, \text{rank}} = \Phi_t^Tw_t^{\epsilon, \text{rank}},
\end{equation}
As in name space, any $w_t^{\epsilon, \text{rank}}$ generates a market-neutral portfolio $w_t^{R, \text{rank}}$ (proof in Appendix~\ref{section: market_neutrality}).

\subsection{Trading signals and portfolio weights}\label{seq: trading_signals_portfolio_weights}
After computing the residual returns $\epsilon_t$\footnote{For simplicity, we take a unified notation $\epsilon_t$ to denote the residual returns in both name and rank space in the following discussions unless otherwise specified.}, we derive the cumulative residual returns over a look-back window of length $L$:
\begin{equation}\label{eq: def_cummulative_residual_return}
    x_t^L = (x_{t-L+1}, x_{t-L+2}, ..., x_{t}),
\end{equation}
where $x_{t-L+\alpha} = \sum_{j=1}^{\alpha} \epsilon_{t-L+j}, \quad \alpha=1, 2, \dots, L$. 

We adopt DNNs $\mathcal{N}: x_t^L\rightarrow w_t^{\epsilon|\text{NN, name/rank}}$ as a data-driven method to calculate portfolio weights in residual space $w_t^\epsilon$ for both name space and rank spaces. The DNNs consist of convolutional layers to capture local patterns, followed by transformer encoder layers to model global dependencies. The neural networks are trained via mean-variance optimization,
\begin{equation}\label{eq: portfolio_weights_epsilon_NN_name}
\begin{aligned}
    \text{Maximize}_{\mathcal{N}(\cdot)}\quad &\mathbb{E}[(w_t^{R|\text{NN}, \text{name/rank}})^T(r_{t+1}-r_f)] - \gamma \text{Var}[(w_t^{R|\text{NN}, \text{name}})^T (r_{t+1}-r_f)]\\
    \text{s.t. }\quad & w_t^{R|\text{NN}, \text{name/rank}} = \frac{\Phi_t^T w_t^{\epsilon|\text{NN}, \text{name/rank}}}{||\Phi_t^T w_t^{\epsilon|\text{NN}, \text{name/rank}}||_1} \\
    & w_t^{\epsilon|\text{NN}, \text{name/rank}} = \mathcal{N}(x_t^L),
\end{aligned}
\end{equation}
where $\gamma$ is the risk-aversion factor. We show a schematic for our DNNs architecture in \autoref{fig: schematic_NN_architecture} with architecture and implementation details in Appendix \ref{sec: dnn_details}. 

\begin{figure}
    \centering
    \includegraphics[scale=0.25]{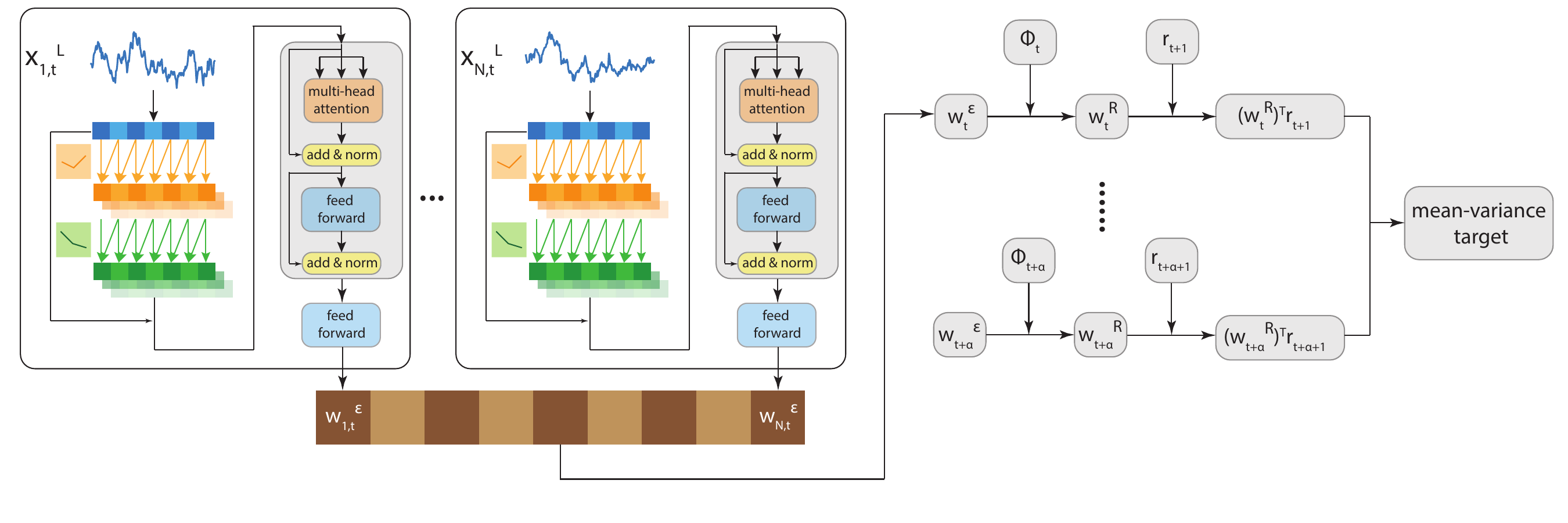}
    \caption{Schematic for the architecture of deep neural networks in both name and rank space.}
    \label{fig: schematic_NN_architecture}
\end{figure}

For comparison, we benchmark the DNN performance against a classical parametric model based on Ornstein-Uhlenbeck (OU) process\cite{avellaneda2010statistical, yeo2017risk}, with the model formulation and execution details in Appendix \ref{sec: ou_process}. 

\subsection{Intraday rebalancing}\label{section: intraday_rebalancing_main_text}
The portfolio weights calculated in the rank space are assigned to artificial financial instruments that yield rank returns in continuous-time limits defined in \eqref{eq: def_return_rank}. To make the constructed portfolio practically implementable, it is necessary to convert these portfolio weights into stock-based portfolios in name space. To address the issue, we propose an intraday rebalancing mechanism. 

Formally, given the predetermined portfolio weights in rank equity space $\{w_{(k),t}^{ \text{rank}}\}_{k=1}^N$ before the market opening, our goal is to rebalance the portfolio at fixed time intervals of $\mathcal{T}$ minutes such that the portfolio becomes $\{w_{(k),t}^{\text{rank}}(1+\Tilde{r}_{(k),t+1})\}_{k=1}^N$ by market close, at the sacrifice of additional costs. To facilitate this discussion, we introduce two processes:

\textbf{(i)} $w_{(k),t+\tau}^{\text{rank}}$: the dollar-valued portfolio weight for rank $k$ at time $t+\tau$;

\textbf{(ii)} $w_{i,t+\tau}^{\text{name}}$: the dollar-valued portfolio weight for stock $i$ at time $t+\tau$.

Here, $t$ denotes the daily time tick and $\tau$ represents the intraday time tick. For instance, for $t=$ Jan, 3rd, 2022 (end of the day) and $\tau=45$ minutes, $t+\tau$ refers to Jan, 4th, 2022 00:45 AM, and $t+1$ refers to the end of the day on Jan, 4th, 2022. 

$w_{(k),t+\tau}^{\text{rank}}$ is the portfolio weight on the $k$-th rank that evolves strictly based on the rank returns in continuous-time limit, 
\begin{equation}\label{eq: intraday_portfolio_weights_rank_dynamics}
    w_{(k),t+\tau}^{\text{rank}} = w_{(k), t}^{\text{rank}}(1+\Tilde{r}_{(k),t+\tau}),
\end{equation}
where $\Tilde{r}_{(k),t+\tau} = \frac{c_{(k), t+\tau}}{c_{(k), t}} - 1$. 
Here, $c_{(k), t+\tau}$ denotes the capitalization at $k$-th rank at time $t+\tau$. 

In contrast, $w_{i,t+\tau}^{\text{name}}$ is the portfolio weight on the $i$-th stock, evolving according to the following rules:

\textbf{(i)} Between the rebalancing interval when $t+j\mathcal{T} < t+\tau \leq t+(j+1)\mathcal{T}, j\in\mathbb{N}$, 
\begin{equation}\label{eq: intraday_portfolio_weights_name_dynamics_1}
    w_{i,t+\tau}^{\text{name}} = w_{i,t+(j\mathcal{T})^+}^{\text{name}}\times\frac{c_{i, t+\tau}}{c_{i,t+(j\mathcal{T})^+}},
\end{equation}where $(j\mathcal{T})^+ := \lim_{\delta \downarrow 0} (j\mathcal{T}+\delta)$.

\textbf{(ii)} At the re-balancing points when $\tau = ((j+1)\mathcal{T})^+, j\in \mathbb{N}$, adjust the portfolio weights via active trading such that
\begin{equation}\label{eq: intraday_portfolio_weights_name_dynamics_2}
    w_{i,t+((j+1)\mathcal{T})^+}^{\text{name}} = \sum_{k=1}^n w_{(k),t+(j+1)\mathcal{T}}^{\text{rank}}\boldsymbol{1}_{\{\mathcal{R}_{i,t+(j+1)\mathcal{T}}= k\}}.
\end{equation}
In other words, we carry out the conversion of portfolio weights between name space and rank space at the rebalancing points $\tau = ((j+1)\mathcal{T})^+, j\in \mathbb{N}$ through active trading. Notably, the value on the trading book before trading at $t+(j+1)\mathcal{T}$ is $\sum_{k=1}^N w_{i,t+(j+1)\mathcal{T}}^{\text{name}}$, while the desired value immediately after trading is $\sum_{i=1}^N w_{i,t+(j+1)\mathcal{T}^+}^{\text{name}} = \sum_{k=1}^N w_{(k),t+(j+1)\mathcal{T}}^{\text{rank}}$. The two values are not necessarily equal when there exists switching of ranks in capitalization between $t+j\mathcal{T}$ and $t+(j+1)\mathcal{T}$ (elaborated by a case study in Appendix \ref{section: intraday_rebalancing} and by the hybrid-Atlas model in Appendix \ref{sec: hybrid-atlas model}). Consequently, the cost of the active trading at $t+(j+1)\mathcal{T}$ involves two components,
\begin{equation}\label{eq: cost_intraday_rebalance}
\begin{aligned}
\text{cost}(t+(j+1)\mathcal{T}^+; w_{(k), t}^{\text{rank}}) = & (\sum_{i=1}^{N} w_{i, t+(j+1)\mathcal{T}^+}^{\text{name}} - \sum_{k=1}^N w_{(k), t+(j+1)\mathcal{T}}^{\text{rank}})\\
& + \eta\sum_{i=1}^N|w_{i, t+(j+1)\mathcal{T}^+}^{\text{name}} - w_{i, t+(j+1)\mathcal{T}}^{\text{name}}|,
\end{aligned}
\end{equation}
where $\eta$ is the transaction cost factor. Here, the cost at time $t+(j+1)\mathcal{T}^+$ depends on the portfolio weights assigned at the beginning of the trading day, $w_{(k), t}^{\text{rank}}$, because the intraday portfolio weights $w_{i, t+\tau}^{\text{name}}$ and $w_{(k), t+\tau}^{\text{rank}}$ are recursively governed by the system dynamics (\eqref{eq: intraday_portfolio_weights_rank_dynamics}, \eqref{eq: intraday_portfolio_weights_name_dynamics_1}, and \eqref{eq: intraday_portfolio_weights_name_dynamics_2}) from the initial condition $w_{(k), t}^{\text{rank}}$. We highlight this dependence by including the $w_{(k), t}^{\text{rank}}$ as a parameter for the cost in \eqref{eq: cost_intraday_rebalance}. We refer to the first term in \eqref{eq: cost_intraday_rebalance} as latency cost, the second term as cost from the bid-ask spread, with their sum representing the total transaction cost. The terminology are rationalized in the case study in Appendix \ref{section: intraday_rebalancing}.

The precise implementation of the intraday rebalancing is summarized in Appendix \ref{section: intraday_rebalancing} Algorithm \ref{alg: intraday_rebalance} along with a schematic in panel (f3) in \autoref{fig: schematic_formulation_framework}. For our backtesting, we primarily use $\eta=2$ basis points to account for the cost from the bid-ask spread. This setting approximately corresponds to a 5-10 cents bid-ask spread for our investment universe, the top 500 stocks in the U.S. equity market.

\subsection{Backtesting}
We evaluate the portfolio performance by calculating the historical profit-and-loss (PnL) $V_t$ and the Sharpe ratio. 

For portfolios in name space, 
\begin{equation}\label{eq: PnL_name_space}
V_{t+1}  = (1+r_{f,t+1})\times (V_t - \sum_i{\Lambda w_{i,t}V_t}-\text{TC}) + \sum_i \Lambda V_t w_{i,t}(1+r_{i, t+1}),
\end{equation}
where $\Lambda=1$ is the leverage, $r_{f,t+1}$ is the risk-free rate during the trading day $t+1$, and $w_{t}$ are normalized by $l_1$ norm. The transaction cost is given by $\text{TC} = \eta \sum_{i}\Lambda |V_t w_{i,t} - V_{t-1}w_{i,t-1}(1+r_{i,t})|,$ where $\eta$ is the transaction cost factor, set to 2 basis points.

For portfolios in rank space, the PnL evolves as
\begin{equation}\label{eq: PnL_rank_space}
\begin{aligned}
    V_{t+1} = & (V_t - \sum_{k=1}^{N}w_{(k), t}^{R, \text{rank}})(1+r_{f, t+1}) + \sum_{k=1}^{N}\Lambda V_t w_{(k), t}^{R, \text{rank}}(1+\Tilde{r}_{(k), t+1}) \\
    & - \sum_{j: t<t+j\mathcal{T}\leq t+1} \text{cost}(t+j\mathcal{T}^+; \Lambda V_t w_{(k), t}^{R, \text{rank}}),
\end{aligned}
\end{equation}
where the last term accounts for the transaction costs due to intraday rebalancing defined in \eqref{eq: cost_intraday_rebalance}, where we substitute generic portfolio weights $w_{(k), t}^{\text{rank}}$ in \eqref{eq: cost_intraday_rebalance} into specific portfolio weights $\Lambda V_t w_{(k), t}^{\text{R, rank}}$ in \eqref{eq: PnL_rank_space}.

\section{Empirical results for the U.S. equities}
\subsection{Market structure: name space versus rank space}
We begin by comparing the market dynamics in name space and rank space, highlighting two key advantages of rank space: (i) a more structured market dynamics in rank space, and (ii) a more enhanced mean-reverting behavior of residual returns in rank space -- both critical motivations for operating statistical arbitrage in rank space. 

\begin{figure}[h!]
    \centering
    \includegraphics[scale=0.25]{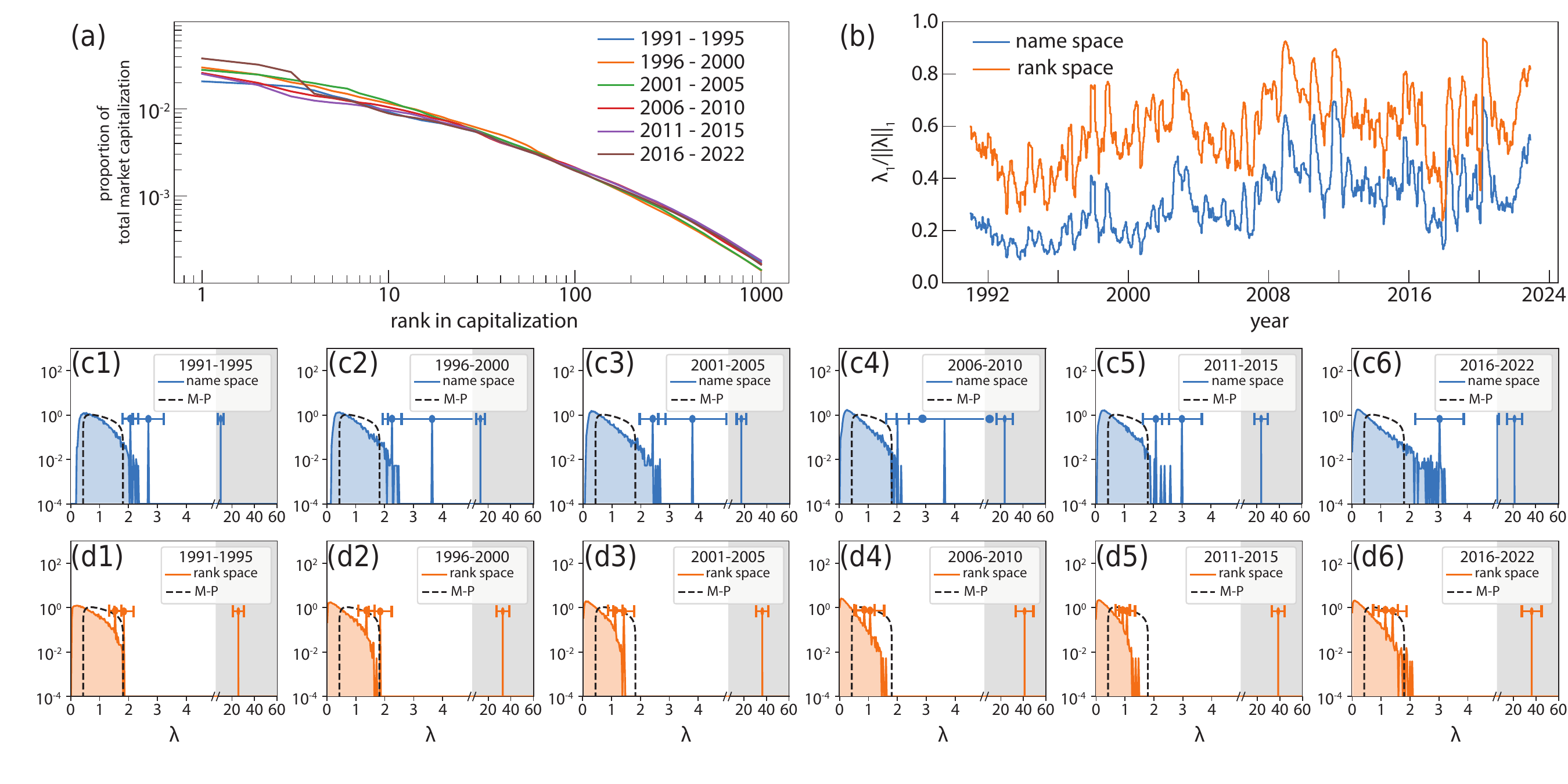}
    \caption{\textbf{Market structure in name space versus rank space. (a)} Proportion of total market capitalization versus ranks in capitalization. \textbf{(b)} The principal eigenvalue of the correlation matrices of $r_t$ in name space (blue) and for $\Tilde{r}_t$ rank space (orange). \textbf{(c, d)} The empirical probability distribution density of the eigenvalue spectrum of the correlation matrices versus Marchenko-Pastur distribution. The market exhibits a more structured correlation in rank space compared to name space.}
    \label{fig: market_structure_name_vs_rank}
\end{figure}

First, we show the market capitalization across ranks, averaged over five-year window from 1991 to 2022 in \autoref{fig: market_structure_name_vs_rank}(a), revealing a stable distribution in rank space\cite{fernholz2002stochastic}. A principal component analysis (PCA) on the correlation matrix suggests significantly larger leading eigenvalue in rank space compared to that in name space (\autoref{fig: market_structure_name_vs_rank}(b)), implying that a greater proportion of market variance is captured by the first eigenvector in rank space. This points to a more structured and concentrated market dynamic in rank space.

More importantly, the single-factor structure in rank space substantially simplifies market decomposition. We present the empirical eigenvalue spectra of the correlation matrix in both spaces across different periods in \autoref{fig: market_structure_name_vs_rank}(c1-c6) and (d1-d6). In name space, several eigenvalues exceed the Marchenko–Pastur upper bound~\cite{avellaneda2022principal}, indicating a multi-factor market and consequently low signal-to-noise ratio(\autoref{fig: market_structure_name_vs_rank}(c1-c6)). In contrast, rank space exhibits a sharp bulk-edge separation with a dominant single factor (\autoref{fig: market_structure_name_vs_rank}(d1-d6)). This clearer structure enables more well-defined separation of market factors from residuals during market decomposition.

\begin{figure}[h!]
    \centering
    \includegraphics[scale=0.25]{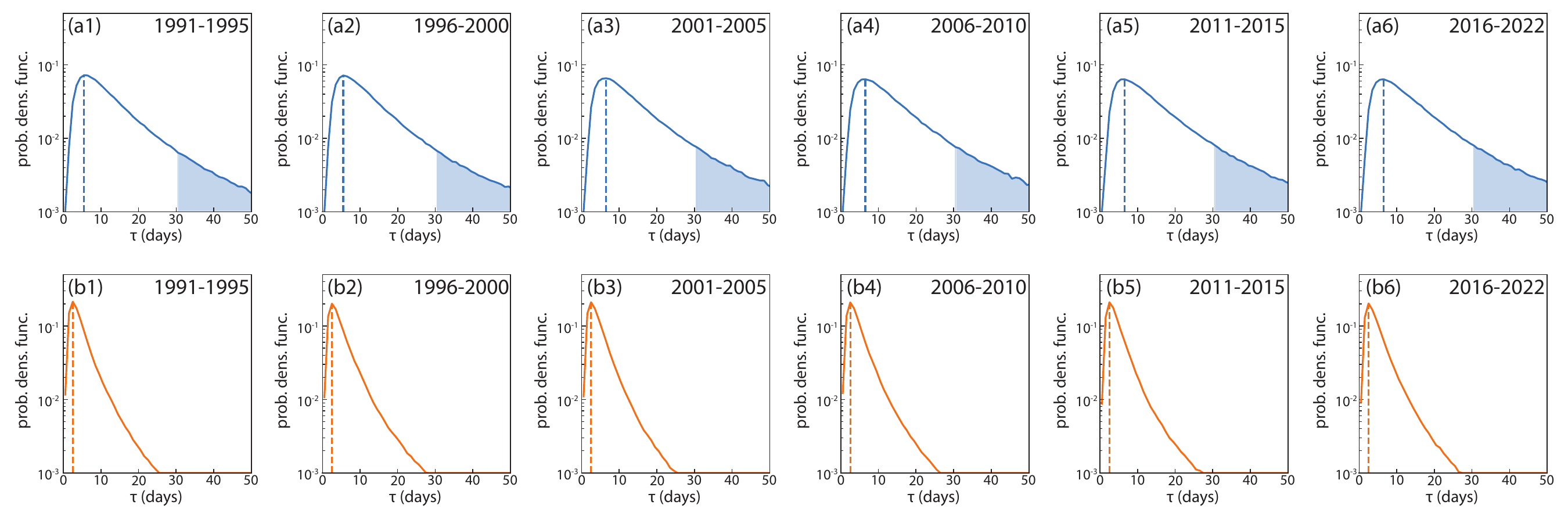}
    \caption{\textbf{Mean-reverting time $\tau$ in name space versus rank space.} The $\tau$ is evaluated by fitting the cumulative residual return $x_t^L$ to an Ornstein–Uhlenbeck(OU) process. \textbf{(a1-a6)} The empirical distributions of mean-reverting time $\tau$ in name space, with maximum empirical probability at $\sim 6$ days (vertical dashed lines). \textbf{(b1-b6)} The empirical distributions of $\tau$ in rank space, with maximum empirical probability at $\sim 2.5$ days (vertical dashed lines). The residual returns in rank space show faster mean-reverting behavior, favorable for statistical arbitrage.}
    \label{fig: mean_reverting_time}
\end{figure}

Second, we observe markedly faster mean-reversion of residual returns in rank space, critical for statistical arbitrage. We quantify mean-reversion by fitting cumulative residual returns $x_t^L$ to an OU process and extracting the mean-reversion time $\tau$. \autoref{fig: mean_reverting_time} shows the empirical distribution of $\tau$ over five-year windows from 1991 to 2022. In name space (\autoref{fig: mean_reverting_time}(a1–a6)), the distribution is heavy-tailed toward large $\tau$, reflecting slower mean reversion. In contrast, rank space (\autoref{fig: mean_reverting_time}(b1–b6)) exhibits a sharper concentration in the fast mean-reverting regime, with fewer instances of slow mean-reversion ($\tau > 30$ days, shaded area).
A non-parametric analysis in Appendix \ref{sec: non_parametric_analysis_on_mean_reversion} yields consistent results.

\subsection{Portfolio performance}
\begin{figure}[h]
    \centering
    \includegraphics[scale=0.25, trim=0 70 0 50, clip]{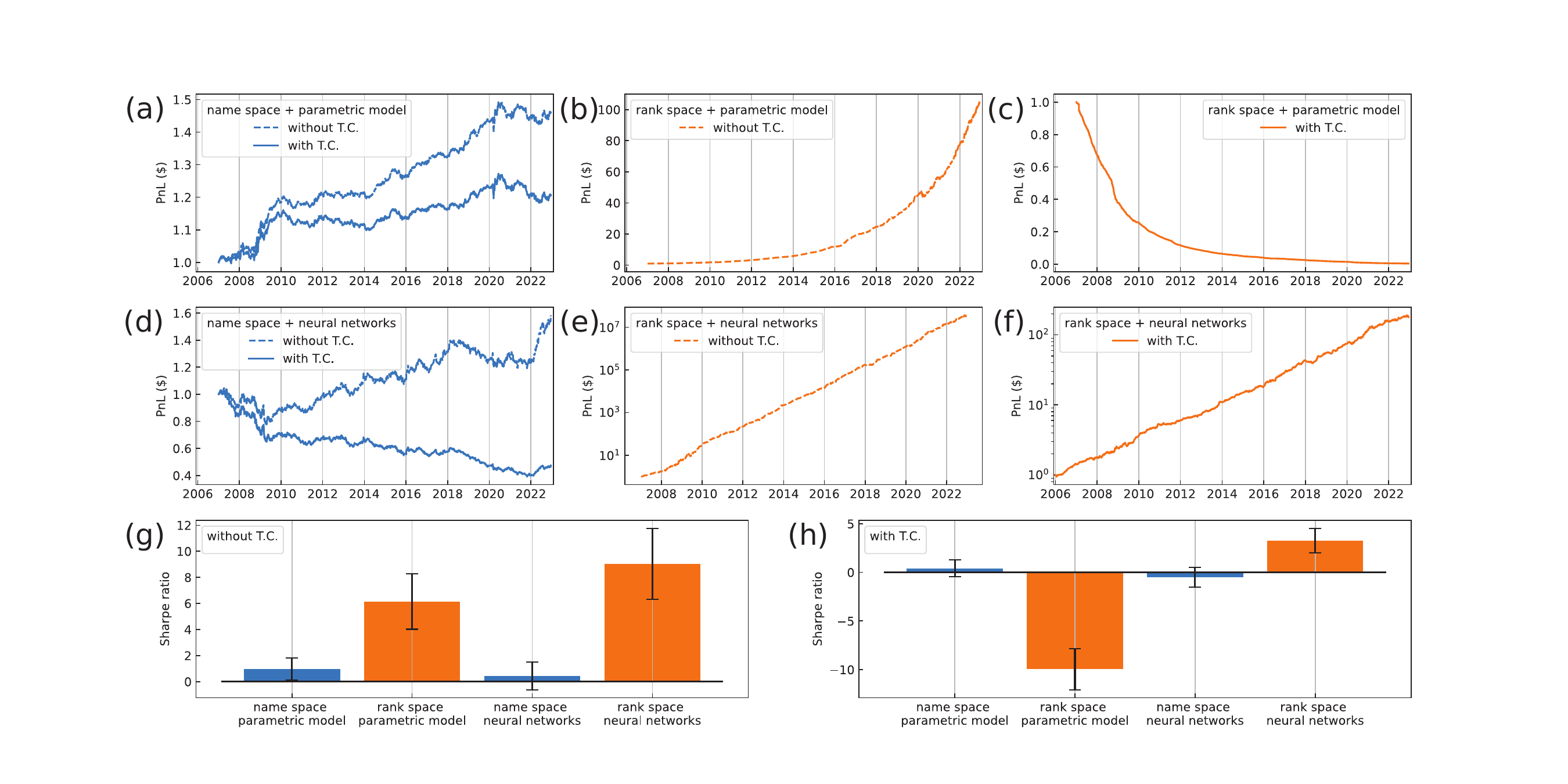}
    \caption{\textbf{Summary of portfolio performance.} The PnL dynamics $V_t$ are computed using \eqref{eq: PnL_name_space} in name space and \eqref{eq: PnL_rank_space} in rank space. 
    \textbf{(a)} PnL using portfolio weights from the parametric model in name space; dashed and solid lines represent results without and with transaction costs, respectively. 
    \textbf{(b, c)} PnL using portfolio weights from the parametric model in rank space; dashed/solid lines represent results without/with transaction costs in panel (b)/(c). 
    \textbf{(d)} Same as (a), but using portfolio weights derived from DNNs in name space. 
    \textbf{(e, f)} Same as (b, c), but using portfolio weights from DNNs in rank space. 
    \textbf{(g, h)} Average Sharpe ratio without/with transaction costs shown in panel (g)/(h). Notably, portfolios derived from DNNs in rank space perform significantly better than those in name space.}
    \label{fig: portfolio_performance_summary}
\end{figure}

We present the PnL $V_t$ in \autoref{fig: portfolio_performance_summary} with portfolio weights $w_t^R$ are calculated by four scenarios: (i) the parametric benchmark model in name space in panel (a), (ii) the parametric benchmark model in rank space in panels (b,c), (iii) DNNs in name space in panel (d), and (iv) DNNs in rank space in panels (e, f). The corresponding Sharpe ratios from 2016 to 2022 are summarized in \autoref{fig: portfolio_performance_summary}(g, h), with year-by-year statistics provided in Appendix~\ref{sec: portfolio_performance_sharp} (without transaction costs: \autoref{table: portfolio_performance_wo_tc}; with transaction costs: \autoref{table: portfolio_performance_with_tc}).

The traditional statistical arbitrage strategy in name space using the parametric model exhibits diminishing profitability after the 2010s. In rank space, the parametric model yields mixed results: while initial performance without transaction costs appears attractive (\autoref{fig: portfolio_performance_summary}(b)), accounting for transaction costs ($\eta=0.0002$) leads to a monotonic decline in PnL (\autoref{fig: portfolio_performance_summary}(c)). This stark contrast highlights a trade-off for statistical arbitrage in rank space: although the residual returns exhibit strong mean reversion and high profit potential, replicating these returns in rank space requires frequent intra-day rebalancing, which incurs substantial transaction costs.

To address these challenges, we leverage DNNs to better exploit market patterns, particularly in rank space. The performance of DNNs in name and rank spaces diverges sharply. In name space, DNNs fail to improve returns or Sharpe ratios (\autoref{fig: portfolio_performance_summary}(a, d); Appendix~\ref{sec: portfolio_performance_sharp}, \autoref{table: portfolio_performance_wo_tc}, \autoref{table: portfolio_performance_with_tc}). In contrast, in rank space, DNNs substantially enhance portfolio performance, achieving an average annual return of 35.68\% and an average Sharpe ratio of 3.28 from 2007 to 2022 (\autoref{fig: portfolio_performance_summary}(h); \autoref{table: portfolio_performance_with_tc}), even after accounting for transaction costs.
This success is driven by the effective exploitation of mean-reversion behavior in rank space by DNNs that yields an average annual return of 206.49\% and an average annual Sharpe ratio of 9.04 without transaction costs (Appendix \ref{sec: portfolio_performance_sharp}, \autoref{fig: portfolio_performance_summary}(g), \autoref{table: portfolio_performance_wo_tc}), sufficient to offset the substantial costs in intraday rebalancing to realize $\Tilde{r}_t$.

Further characterization of the portfolios is provided in Appendix \ref{sec: portfolio_performance}. Appendix \ref{sec: portfolio_performance_sharp} presents annualized performance statistics, Appendix \ref{sec: portfolio_performance_market_dollar_neutrality} examines market and dollar neutrality, Appendix \ref{sec: portfolio_performance_transaction_costs} discusses dependence on transaction costs, and Appendix \ref{sec: characteristic_time_rank_swap} analyzes the role of rank-swapping timescales on portfolio performance.

\subsection{The intelligence inside the neural networks}
To understand the outperformance of DNNs compared to the benchmark parametric model in rank space, we analyze the relationship between the input (the trajectories of the cumulative residual return $x_t^L$) and the output (the portfolio weights in residual space $w_t^\epsilon$). 

\begin{figure}[h!]
    \centering
    \includegraphics[scale=0.25]{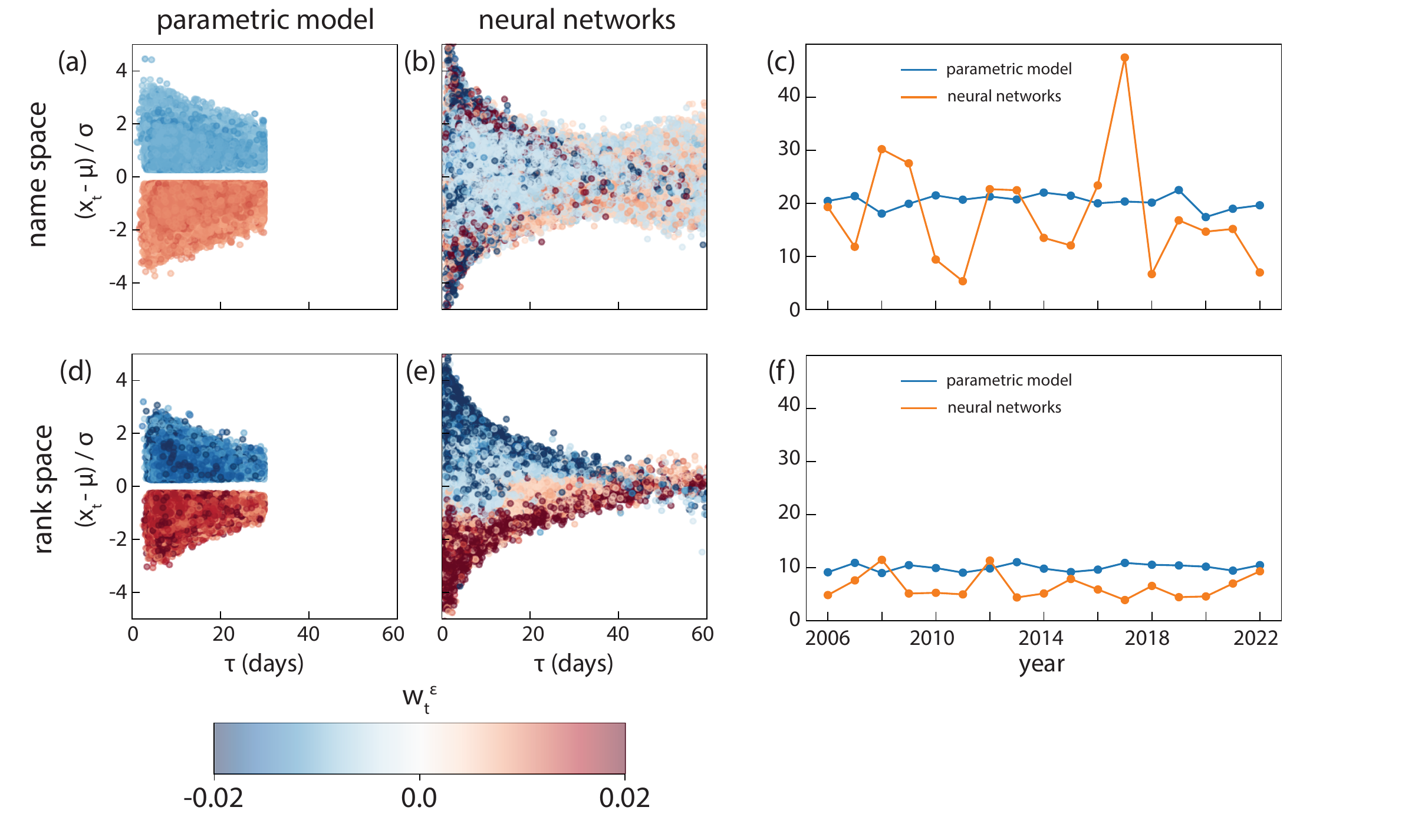}
    \caption{\textbf{Portfolio weights in residual space: parametric model versus neural networks.} \textbf{(a, b, d, e)} We illustrate the behavior of both the parametric model and neural networks by analyzing the relationship between the input, the trajectories of cumulative residual returns $x_t^L$, and the output, the portfolio weights in residual space, $w_t^\epsilon$. The input $x_t^L$ are parameterized by two variables: (i) deviation from long-term average $\frac{x_t-\mu}{\sigma}$, and (ii) mean-reverting time $\tau$. Each $x_t^L$ thus corresponds to a point in the plane spanned by $\frac{x_t-\mu}{\sigma}$ and $\tau$, color-coded by $w_t^\epsilon$. This analysis is performed with $w_t^\epsilon$ calculated by four scenarios: (a) the parametric model in name space; (b) neural networks in name space; (d) the parametric model in rank space; (e) neural networks in rank space. \textbf{(c, f)} Panel (c) and (f) show the average holding time for portfolios derived in name space and rank space, respectively. The DNN-derived portfolios exhibit more intelligence in terms of variable leverage, flexible opportunity thresholds, and shorter holding periods.}
    \label{fig: interprete_neural_network}
\end{figure}

We parameterize $x_t^L$ by two key variables: the deviation from long-term average $\frac{x_t-\mu}{\sigma}$, and the mean-reverting time $\tau$, following the trading signal suggested by the parametric model(\eqref{eq: portfolio_weights_epsilon_OU_process}). Each $x_t^L$ thus corresponds to a point in the plane spanned by $\frac{x_t-\mu}{\sigma}$ and $\tau$, color-coded by $w_t^\epsilon$ (\autoref{fig: interprete_neural_network}). 

We evaluate four scenarios: (i) parametric model in name space (\autoref{fig: interprete_neural_network}(a)), (ii) DNNs in name space (\autoref{fig: interprete_neural_network}(b)), (iii) parametric model in rank space (\autoref{fig: interprete_neural_network}(d)), and (iv) DNNs in rank space (\autoref{fig: interprete_neural_network}(e)). We also report the average holding periods before liquidation for each method (\autoref{fig: interprete_neural_network}(c, e)).

Compared to the parametric benchmark, the DNNs demonstrate more sophisticated trading behavior in rank space. Despite operating directly on raw cumulative return trajectories, the DNNs successfully uncover the significance of mean-reversion through mean-variance optimization (\autoref{fig: interprete_neural_network}(e)). In addition, the DNNs improve the execution strategy along three dimensions:

(i) Variable leverage on deviations. The DNNs assign higher leverage to positions with larger normalized deviations $\frac{x_t-\mu}{\sigma}$, enhancing profit potential during significant market moves (\autoref{fig: interprete_neural_network}(e)).

(ii) Flexible opportunity thresholds. Rather than relying on rigid mean-reversion time ($\tau$) cutoffs as in the parametric model (\autoref{fig: interprete_neural_network}(a, d)), DNNs embrace a broader range of trading opportunities while concentrating investments in trajectories with fast mean-reversion ($\tau$ small) (\autoref{fig: interprete_neural_network}(b, e)).

(iii) Shorter holding periods. DNNs reduce average holding periods to around 5 days, compared to approximately 10 days under the parametric model (\autoref{fig: interprete_neural_network}(c, e)).
This minimizes carry-over risk, which is particularly important when employing variable leverage across positions.

We also emphasize the critical role of market data preprocessing.
While the DNNs achieve substantial improvements in rank space, they fail to converge or deliver gains in name space (\autoref{fig: training_curve_NN}; \autoref{fig: portfolio_performance_summary}(d)).
Despite both input spaces being derived from similar capitalization data, strategic reorganization into rank space dramatically improves training efficiency (\autoref{fig: training_curve_NN}) and portfolio performance (\autoref{fig: portfolio_performance_summary}). This illustrates the importance of domain-informed data transformations in financial machine learning: appropriate restructuring of the input space can substantially enhance the learning efficiency and performance of deep models in complex, noisy environments like equity markets.

\section{Limitations and discussion}
Our current strategy is sensitive to transaction costs due to the frequent intraday rebalancing required to realize return in rank space, as it ceases to profit with 5 bps transaction costs (Appendix~\ref{sec: portfolio_performance_transaction_costs}). The induced transaction costs and slippage may limit the achievable volume and scalability in practice. Furthermore, compared to a traditional parametric model in name space, portfolio weights derived from neural networks exhibit significantly higher volatility (Figure \ref{fig: dollar_neutrality} in Appendix~\ref{sec: portfolio_performance_market_dollar_neutrality}), raising practical concerns for risk managements.

While our proposed intraday rebalancing mechanism (Appendix~\ref{section: intraday_rebalancing}) provides a baseline, it offers substantial room for further optimization. Future work could formulate this challenge as a stochastic control problem, leveraging physics-informed neural networks~\cite{raissi2019physics} to solve the resulting high-dimensional partial differential equations, or alternatively as a reinforcement learning problem based on real-market data. 

The advantage of statistical arbitrage in rank space relies on frequent switching of ranks in capitalization, resulting in a qualitatively different return space compared to that in name space. Therefore, we expect that the advantage may be relevant to other volatile financial markets. Beyond financial markets, the rank-based representation introduced here may generalize to other many-particle systems, such as those encountered in many-particle physics\cite{mahan2013many}, biology\cite{berg1993random}, and social sciences\cite{garcia2013sociophysics}.

\section{Conclusion}
We have introduced a novel statistical arbitrage method that leverages the robust market structure in rank space. Although rank space and name space contain the same underlying information, the significant performance improvements achieved by DNNs in rank space highlight the critical importance of domain-informed data representations in financial machine learning. In particular, our results demonstrate how appropriate transformations of the input space can dramatically enhance the learning efficiency and performance of deep models in complex, noisy environments such as many-particle systems like equity markets.

\clearpage
\appendix
\section{Implementation details}\label{sec: implementation_details}
\subsection{Data and experimental setup}
We collect dividend-adjusted daily return, price, shares outstanding, and capitalizations for the U.S. securities from Center for Research in Security Prices (CRSP), covering January 1990 to December 2022. Intraday price data at 1-minute resolution from January 2005 to December 2022 are obtained from \textit{Polygon.io}. We construct intraday capitalization data by combining CRSP shares outstanding with \textit{Polygon.io} intraday prices. The one-month Treasury bill rate from the Kenneth French Data Library is used as the risk-free rate $r_f$.

\subsection{Calibrating investment universe and market decomposition}\label{sec: backtesting_details}
Our backtesting spans January 2006 to December 2022, covering both the subprime mortgage crisis and the highly competitive post-2010 market period.
On each trading day after market close, we re-calibrate the investment universe by selecting stocks that (i) rank among the top 500 in capitalization as of day $t$, ensuring sufficient liquidity, and (ii) have valid historical return data available for day $t+1$. This selection procedure minimizes potential selection bias to the best of our ability.

We then perform principal component analysis (PCA) on the selected returns using a 252-day lookback window to extract leading eigenvectors as market factors $F_t$. Specifically, we retain the top five eigenvectors (associated with the five largest eigenvalues) in name space, and the top eigenvector in rank space. Factor loadings $\beta_t$ are estimated using a 60-day lookback window, from which the transformations $\Phi_t$ and residual returns $\epsilon_t$ are computed. Cumulative residual returns $x_t^L$ are similarly evaluated over the same 60-day window and are used as inputs to either the parametric benchmark model (Appendix~\ref{sec: ou_process}) or the deep neural networks to generate portfolio weights and calculate the resulting PnL.

\subsection{Deep neural networks}\label{sec: dnn_details}
We delve into the specific architecture of our neural networks, illustrated in \autoref{fig: schematic_NN_architecture}. Our CNN-transformer architecture harnesses the strengths of CNN in extracting local patterns and transformers in capturing long-term dependencies. The inputs of our neural networks are the trajectories of cumulative residual returns, $x_t^L\in\mathbb{R}^{N\times L}$, processed through two layer of multi-channel convolutional networks, followed by a standard transformer encoder layer that models global relationships via multi-head attention. Specifically, in the convolutionary layer,
\begin{equation}
\begin{aligned}
    x_t^{(1)} &= \frac{x_t^L-\mathbb{E}(x_t^L)}{\sqrt{\text{Var}(x_t^L)+\epsilon}}\times\gamma^{(1)}+\beta^{(1)}, \quad y_t^{(1)} = W^{(1)}*x_t^{(1)} + b^{(1)},\quad z_t^{(1)} = \text{ReLu}(y_t^{(1)}) + x_t^{(1)};\\
    x_t^{(2)} & = \frac{z_t^{(1)}-\mathbb{E}(z_t^{(1)})}{\sqrt{\text{Var}(z_t^{(1)})+\epsilon}}\times\gamma^{(2)}+\beta^{(2)}, \quad y_t^{(2)} = W^{(2)}* x_t^{(2)} + b^{(2)},\quad z_t^{(2)} = \text{ReLu}(y_t^{(2)}) + x_t^{(2)}  .
\end{aligned}
\end{equation}
The superscript $(1)$ or $(2)$ specifies the layer number. $x_t^{(1)}\in \mathbb{R}^{N\times L}$ is the input of the first convolutional layer. $W^{(1)}\in\mathbb{R}^{D_{\text{channel}}\times 1\times D_{\text{kernel}}}$ and $W^{(2)}\in\mathbb{R}^{D_{\text{channel}}\times D_{\text{channel}}\times D_{\text{kernel}}}$ are the convolutionary kernels for the convolution operator denoted by $*$, $b^{(1,2)}\in \mathbb{R}^{D_{\text{channel}}}$ is the bias, and $y_t^{(1,2)}\in\mathbb{R}^{N\times D_{\text{channel}}\times L}$ is the output of convolutionary operator. $D_{\text{channel}}$ is the number of channels and $D_{\text{kernel}}$ is the size of the convolution kernel. We adopt a rectified linear unit (denoted as ReLu($\cdot$)) as our activation function. We also apply (i) instance normalization \cite{ulyanov2016instance} with learnable parameter $\gamma^{(1, 2)}$ and $\beta^{(1, 2)}$ at the input of each convolution layer to accelerate the training process, and (ii) residual connection\cite{he2016deep} to avoid vanishing gradients by directly connecting the input $x_t^{(1,2)}$ to the output $z_t^{(1,2)}\in\mathbb{R}^{N\times D_{\text{channel}}\times T}$. We choose the hyper-parameters for our neural networks as number of channels $D_{\text{channel}}=8$ and size of the convolution kernel $D_{\text{kernel}}=2$.

The outputs of convolutionary layers, $z_t^{(2)}\in\mathbb{R}^{N\times D_{\text{channel}}\times T}$ are subsequently fed into a standard transformer encoder layer\cite{vaswani2017attention}. The transformer encoder layer utilizes the multi-head attention modeled by the inner product between the famous key-query-value matrices. To elaborate, 
\begin{equation}
\begin{aligned}
&x_t^{\text{transformer}} = (z_t^{(2)})^T\\
&\begin{cases}
    Q_i &= \text{DropOut}(W_i^Qx_t^{\text{transformer}} + b_i^Q)\\
    K_i &= \text{DropOut}(W_i^Kx_t^{\text{transformer}} + b_i^K)\\
    V_i &= \text{DropOut}(W_i^Vx_t^{\text{transformer}} + b_i^V)
\end{cases}, \quad i=1,2,..., H\\
&\text{head}_i = \text{softmax}(\frac{Q_iK_i^T}{\sqrt{d_{\text{channel}}/H}}),\quad i=1,2,..., H\\
& y_t = \text{Concat}(\text{head}_1V_1, ..., \text{head}_HV_H)\\
& z_t = \text{LayerNorm}(x_t^{\text{transformer}}+y_t)\\
&o_t = \text{LayerNorm}(W^Oz_t + b^O+y_t),
\end{aligned}
\end{equation}
where $x_t^{\text{transformer}}\in \mathbb{R}^{N\times T\times D_{\text{channel}}}$ is the input of the transformer encoder layer. In addition, $W_i^Q\in\mathbb{R}^{(D_{\text{channel}}/H)\times D_{\text{channel}}}$, $W_i^K\in\mathbb{R}^{(D_{\text{channel}}/H)\times D_{\text{channel}}}$, and $W_i^V\in\mathbb{R}^{D_{(\text{channel}}/H)\times D_{\text{channel}}}$ are the linear weights. $b_i^Q\in\mathbb{R}^{D_{\text{channel}}/H}$, $b_i^K\in\mathbb{R}^{D_{\text{channel}}/H}$, and $b_i^V\in\mathbb{R}^{D_{\text{channel}}/H}$ are the bias. Softmax($\cdot$) stands for softmax function and Concat($\cdot$) for matrix concatenation, and $y_t\in\mathbb{R}^{N\times T\times D_{\text{channel}}}$ is the output of multi-attention layer. $W^O\in\mathbb{R}^{D_{\text{channel}}\times D_{\text{channel}}}$ and $b^O\in\mathbb{R}^{D_{\text{channel}}}$ are the linear weights and bias in the output linear layer. $o_t\in \mathbb{R}^{N\times T\times D_{\text{channel}}}$ is the output of the transformer. In addition to the residual connection similar to the convolutional layer, we also introduce the drop-out technique, denoted as Dropout($\cdot$), to regularize overfitting with drop-out probability $p$, and layer normalization\cite{ba2016layer}, denoted as LayerNorm($\cdot$), to improve training stability. We choose the hyper-parameters for our neural networks as the number of heads $H=4$ and the drop-out probability $p=0.25$.

Finally, we choose the last slice along the time axis in the output of the transformer, $o_t\in \mathbb{R}^{N\times T\times D_{\text{channel}}}$, as the hidden state summarizing the information up to time $t$. The portfolio weights in residual space $w_t^{\epsilon|\text{NN}, \text{name/rank}}$ are calculated by a linear relationship,
\begin{equation}
w_t^{\epsilon|\text{NN}, \text{name/rank}} = W^F(o_t[\text{:, -1, :}])+b^F,
\end{equation}
where $o_t[\text{:, -1, :}]\in\mathbb{R}^{N\times D_{\text{channel}}}$ means the last slice along the second-dimension (time-axis) of $o_t$, and $W^F\in\mathbb{R}^{1\times D_{\text{channel}}}$, $b^F\in\mathbb{R}$ are the parameters.

For the mean-variance optimization target in \eqref{eq: portfolio_weights_epsilon_NN_name}, the empirical expectation and variance are obtained over a consecutive time window of length $T$,
\begin{equation}
\begin{aligned}
\mathbb{E}[(w_t^{R|\text{NN}})^T (r_{t+1}-r_f)] &\approx \frac{1}{T}\sum_{\alpha=1}^T (w_{t+\alpha}^{R|\text{NN}})^T (r_{t+\alpha+1}-r_f)\\
\text{Var}[(w_t^{R|\text{NN}})^T (r_{t+1}-r_f)] &\approx \frac{1}{T}\sum_{\alpha=1}^T [(w_{t+\alpha}^{R|\text{NN}})^T (r_{t+\alpha+1}-r_f) - \mathbb{E}((w_t^{R|\text{NN}})^T (r_{t+1}-r_f))]^2
\end{aligned}
\end{equation}
We choose the risk-aversion factor $\gamma=2$ and length of time window $T=24$ days.

\begin{figure}[h]
    \centering
    \includegraphics[width=\columnwidth]{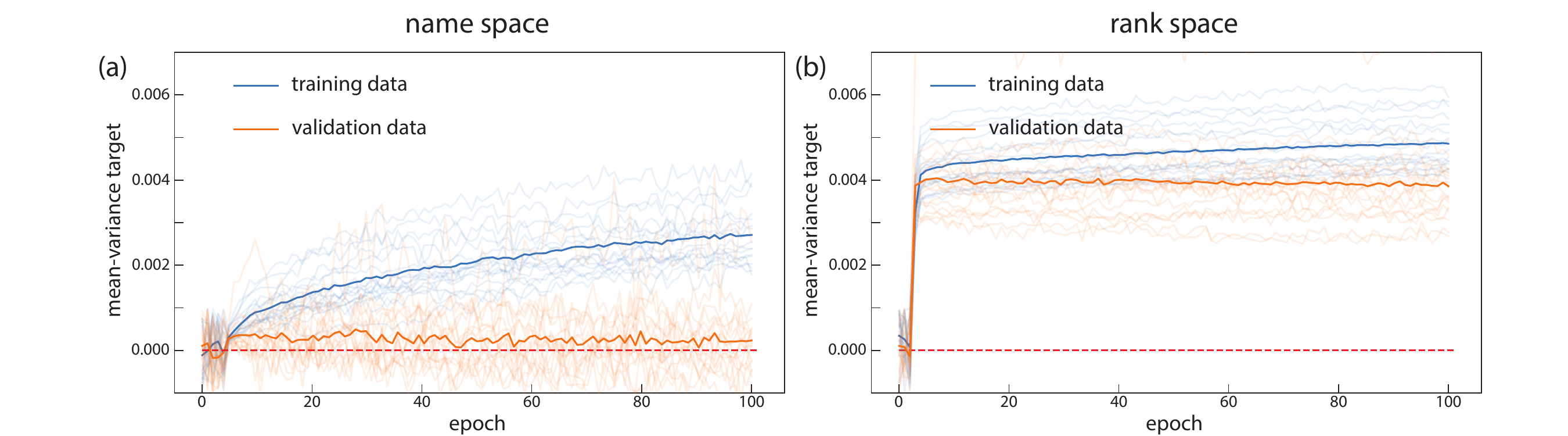}
    \caption{Training curves of neural networks. \textbf{(a, b)} We show the mean-variance target as a function of training epochs for neural networks as specified by \eqref{eq: portfolio_weights_epsilon_NN_name} in name space (a) and in rank space (b). The training curves originate from the phase I training process focusing on hyperparameter tuning. To evaluate the out-of-sample portfolio weights from trading day $t$ to $t+252$, we use the training data from day $t-1000$ to day $t-60$ and validation data from day $t-59$ to $t-1$ as the validation data. The neural networks are re-trained annually with random initialization, yielding approximately 17 training curves from 2006 to 2022. The transparent lines show individual training curves with their average represented by the opaque lines. The neural networks in rank space are more efficiently trained than those in name space.}
    \label{fig: training_curve_NN}
\end{figure}

The neural networks are trained in two steps. The first step aims at optimizing the hyper-parameters of neural network. Specifically, to evaluate the portfolio weights from the trading day $t$ to day $t+252$, we utilize the data from $t-1000$ to $t-60$ as the training data set and from $t-59$ to $t-1$ as the validation data set, from which we determine hyper-parameters from the converged mean-variance target in the training curve (\autoref{fig: training_curve_NN}). The hyperparameters of our optimized neural network architecture are elaborated in Appendix \ref{sec: dnn_details}.

The second step aims at increasing the updating frequency of parameters in the neural networks from annually to quarterly while keeping hyper-parameters fixed. More explicitly, to evaluate the portfolio weights from the trading day $t$ to day $t+63$, we use the data from trading day $t-500$ to day $t-1$ as the training data set. Empirically, increasing the updating frequency is significantly beneficial to the performance of neural network, due to the non-stationarity in financial data. The training tasks utilize PyTorch 2.2.0, and are parallelized on a workstation with a CPU from AMD Ryzen Threadripper Pro 5955 WX and two GPUs from Nvidia GeForce RTX 4090. Each neural network training iteration takes approximately two hours to converge, resulting in roughly 130 computational hours for backtesting portfolio performance from 2007 to 2022, assuming quarterly neural network retraining. In our rank-space statistical arbitrage strategy, the daily portfolio weights in rank space are precomputed by forward propagation of the neural network before market opening, and remain fixed during the trading day. To handle rank changes, the intraday rebalancing of portfolio weights from rank space to name space occurs every 225 minutes.

\clearpage
\subsection{Algorithms}\label{section: algorithms}
We provide implementation details in the form of pseudocode. Algorithm \ref{alg: market_decomposition} performs market decomposition in both name and rank space, as discussed in section \ref{section: market_decomposition}. Algorithm \ref{alg: portfolio_weights_parametric_model} calculates the portfolio weights using the parametric model (section \ref{sec: ou_process}), while algorithm \ref{alg: portfolio_weights_neural_network} does so via neural networks (section \ref{seq: trading_signals_portfolio_weights} and \ref{sec: dnn_details}). Algorithm \ref{alg: intraday_rebalance} handles the conversion of portfolio weights between name space and rank space (section \ref{section: intraday_rebalancing}). Algorithm \ref{alg: portfolio_metric} computes the summary statistics of portfolio performance reported in \ref{table: portfolio_performance_wo_tc} and \ref{table: portfolio_performance_with_tc}. Finally, algorithm \ref{alg: stats_arb_name_space} and \ref{alg: stats_arb_rank_space} integrate the proposed algorithmic components above to implement the complete statistical arbitrage strategy in name and rank space, respectively.

\begin{algorithm}[h!]
\caption{Market decomposition (PCA) [\autoref{fig: schematic_formulation_framework}, panel(c1, c2)]}\label{alg: market_decomposition} 
\begin{algorithmic}
\State \textbf{Input}: $r_t, r_{f,t}, K$ 
\State \textbf{Output}: $\epsilon_t, \Phi_t$\\
\State \textbf{Function} market\_decomposition($r_t, r_{f,t}, K$):
\State \hspace{\algorithmicindent} Principal component analysis: $r_t-r_{f,t} = U\Sigma V^T$
\State \hspace{\algorithmicindent} $F_t \leftarrow (v_1, v_2, ..., v_K)$, where $v_k$ is the $k$-th column of $V^T$
\State \hspace{\algorithmicindent} Calculate $\omega_t$ by solving $F_t = \omega_t (r_t - r_f)$
\State \hspace{\algorithmicindent} Calculate $\beta_t$ as the coefficient of the linear regression $r_t - r_f \sim F_t$
\State \hspace{\algorithmicindent} $\Phi_t \leftarrow I - \beta_t \omega_t$
\State \hspace{\algorithmicindent} $\epsilon_t \leftarrow \Phi_t (r_t-r_{f,t})$
\State \hspace{\algorithmicindent}\textbf{return} $\epsilon_t, \Phi_t$\\
\State \tcp{Input:\\$r_t$: return in name space or transformed return in rank space.}
\State \hspace{\algorithmicindent} \tcp{$r_{f,t}$: risk-free rate at the end of trading day $t$.}
\State \hspace{\algorithmicindent} \tcp{$K$: number of market factors, predetermined by analyzing eigenvalue}
\State \hspace{\algorithmicindent} \hspace{\algorithmicindent} \space \space \tcp{spectrum of the correlation matrix.}
\State \tcp{Output:\\$\epsilon_t$: residual returns in name space or rank space.\\$\Phi_t$: transformation between residual space and equity space}
\State \hspace{\algorithmicindent} \space \space \space \space \space \space \space \space \tcp{(\eqref{eq: market_decomposition_name_main} for name space and \eqref{eq: market_decomposition_PCA_rank_main} for rank space).}
\State \tcp{Note:}
\State \hspace{\algorithmicindent} \tcp{The algorithm realizes the formulation in section 2.1.}
\State \hspace{\algorithmicindent} \tcp{Factors $F_t$ and $\omega_t$ are calculated on a 252-day look-back window.}
\State \hspace{\algorithmicindent} \tcp{Loadings $\beta_t$ are calculated on a 60-day look-back window.}
\State \hspace{\algorithmicindent} \tcp{$F_t$, $\omega_t$, and $\beta_t$ are updated daily.}
\State \hspace{\algorithmicindent} \tcp{$K=5$ for name space and $K=1$ for rank space based on empirical}
\State \hspace{\algorithmicindent} \tcp{eigenvalue spectrum of the correlation matrix (\autoref{fig: market_structure_name_vs_rank}(c,d))).}
\end{algorithmic}
\end{algorithm}

\begin{algorithm}[H]
\caption{Portfolio weights by parametric model [\autoref{fig: schematic_formulation_framework}, panel(d1, e1)]}\label{alg: portfolio_weights_parametric_model}
\begin{algorithmic}
\State \textbf{Input}: $\epsilon_t, \Phi_t$
\State \textbf{Output}: $w_t^{R|\text{OU}}$\\
\State \textbf{Function} portfolio\_weights\_by\_parametric\_model($\epsilon_t, \Phi_t)$
\State \hspace{\algorithmicindent} Calculate $x_t^L$ by \eqref{eq: def_cummulative_residual_return}
\State \hspace{\algorithmicindent} Estimate $\tau, \mu, \sigma, x_t, R^2$ by fitting $x_t^L$ to an OU process (\eqref{eq: def_OU_process})
\State \hspace{\algorithmicindent} Calculate $w_t^{\epsilon|\text{OU}}$ by \eqref{eq: portfolio_weights_epsilon_OU_process}
\State \hspace{\algorithmicindent} $w_t^{R|\text{OU}} \leftarrow \Phi_t^Tw_t^{\epsilon|\text{OU}}$(\eqref{eq: portfolio_weights_equity_space_OU})
\State \hspace{\algorithmicindent} \textbf{return} $w_t^{R|\text{OU}}/||w_t^{R|\text{OU}}||_1$\\
\State \tcp{Input:}
\State \tcp{$\epsilon_t$: residual returns calculated from Algorithm \ref{alg: market_decomposition}}
\State \tcp{$\Phi_t$: transformation matrix between equity space and residual space from}
\State \hspace{\algorithmicindent} \space \space \tcp{Algorithm \ref{alg: market_decomposition}}
\State \tcp{Output:}
\State \tcp{$w_t^{R|\text{OU}}$: $l_1$-normalized portfolio weights by parametric model. }
\State \hspace{\algorithmicindent} \hspace{\algorithmicindent} \tcp{For name space, it stands for the portfolio weights on stocks.}
\State \hspace{\algorithmicindent} \hspace{\algorithmicindent} \tcp{For rank space, it corresponds to the portfolio weights on}
\State \hspace{\algorithmicindent} \hspace{\algorithmicindent} \tcp{artificial financial instruments that realize rank returns}
\State \hspace{\algorithmicindent} \hspace{\algorithmicindent} \tcp{defined in \eqref{eq: def_return_rank}.}
\State \tcp{Note:}
\State \hspace{\algorithmicindent} \tcp{The algorithm realizes the formulation in section 2.2.1.}
\State \hspace{\algorithmicindent} \tcp{$\tau$, $\mu$, $\sigma$ are fitting parameters of OU process.}
\State \hspace{\algorithmicindent} \tcp{Risk control by ignoring $\tau>30$ days (\eqref{eq: portfolio_weights_epsilon_OU_process}).}
\end{algorithmic}
\end{algorithm}

\begin{algorithm}[H]
\caption{Portfolio weights by neural networks [\autoref{fig: schematic_formulation_framework}, panel(d2, e2)]}\label{alg: portfolio_weights_neural_network}
\begin{algorithmic}
\State \textbf{Input}: $\epsilon_t, \Phi_t$
\State \textbf{Output}: $w_t^{R|\text{NN}}$\\
\State \textbf{Function} portfolio\_weights\_by\_neural\_networks($\epsilon_t, \Phi_t)$
\State \hspace{\algorithmicindent} Calculate $x_t^L$ by \eqref{eq: def_cummulative_residual_return}
\State \hspace{\algorithmicindent} Train neural network in-sample for mean-variance optimization (\eqref{eq: portfolio_weights_epsilon_NN_name})
\State \hspace{\algorithmicindent} Calculate $w_t^{\epsilon|\text{NN}}$ out-of-sample from trained neural network
\State \hspace{\algorithmicindent} \textbf{return} $w_t^{R|\text{NN}}/||w_t^{R|\text{NN}}||_1$\\
\State \tcp{Input:}
\State \tcp{$\epsilon_t$: residual returns calculated from Algorithm \ref{alg: market_decomposition}.}
\State \tcp{$\Phi_t$: transformation matrix between equity space and residual space from}
\State \hspace{\algorithmicindent} \space \space \tcp{Algorithm \ref{alg: market_decomposition}.}
\State \tcp{Output:}
\State \tcp{$w_t^{R|\text{NN}}$: $l_1$-normalized portfolio weights by parametric model.}
\State \hspace{\algorithmicindent} \hspace{\algorithmicindent} \tcp{For name space, it stands for the portfolio weights on stocks.}
\State \hspace{\algorithmicindent} \hspace{\algorithmicindent} \tcp{For rank space, it corresponds to the portfolio weights on}
\State \hspace{\algorithmicindent} \hspace{\algorithmicindent} \tcp{artificial financial instruments that realize rank returns.}
\State \hspace{\algorithmicindent} \hspace{\algorithmicindent} \tcp{defined in \eqref{eq: def_return_rank}.}
\State \tcp{Note:}
\State \hspace{\algorithmicindent} \tcp{The algorithm realizes the formulation in section 2.2.2.}
\State \hspace{\algorithmicindent} \tcp{No pre-screening on trading opportunities $x_t^L$ applied.}
\State \hspace{\algorithmicindent} \tcp{Neural networks internally prioritize various trading opportunities}
\State \hspace{\algorithmicindent} \hspace{\algorithmicindent} \tcp{and manage risk (\autoref{fig: interprete_neural_network}).}
\State \hspace{\algorithmicindent} \tcp{The mean-variance target is evaluated on a 24-day window.}
\end{algorithmic}
\end{algorithm}
\begin{algorithm}[H]
\caption{intraday rebalancing [\autoref{fig: schematic_formulation_framework}, panel (f3)]}\label{alg: intraday_rebalance}
\begin{algorithmic}
\State \textbf{Input}: $w_t^R, r_{f,t}, \mathcal{T}$
\State \hspace{\algorithmicindent} \hspace{\algorithmicindent} $c_{t+\tau}, \quad t=1,2,..., T$ (days), $\tau=1,2,..., N$ (minutes)
\State \textbf{Output}: $V_t, \quad t=1,2, ..., T+1$\\
\State \textbf{Function} intraday\_rebalancing($w_t^R, r_{f,t}, \mathcal{T}, c_{t+\tau})$
\State \hspace{\algorithmicindent} $t\leftarrow 0$, $V_t\leftarrow 1$, $w^{\text{prev}}\leftarrow 0$
\State \hspace{\algorithmicindent} \textbf{While} $t\leq T$ \Comment{$T$ is daily time tick} 
\State \hspace{\algorithmicindent} \hspace{\algorithmicindent} $w_{(k), t}^{\text{rank}} \leftarrow w_{(k), t}^R, \quad k=1,2,..., N$
\State \hspace{\algorithmicindent} \hspace{\algorithmicindent} $w_{\mathcal{I}_{(k), t}, t}^{\text{name}} \leftarrow w_{(k), t}^R, \quad k=1,2,..., N$ \Comment{$\mathcal{I}_{(k), t}$ maps from rank to name}
\State \hspace{\algorithmicindent} \hspace{\algorithmicindent} $V_t \leftarrow V_t - \sum_{i}w_{i,t}^{\text{name}} - 0.0002\times ||w_t^{\text{name}}-w^{\text{prev}}||_1$
\State \hspace{\algorithmicindent} \hspace{\algorithmicindent} $\tau\leftarrow 0$
\State \hspace{\algorithmicindent} \hspace{\algorithmicindent} \textbf{While} $t+\tau<(t+1)$ \Comment{$\tau$ is intraday time tick}
\State \hspace{\algorithmicindent} \hspace{\algorithmicindent} \hspace{\algorithmicindent} $w_{(k), t+\tau}^{\text{rank}} \leftarrow w_{(k), t+\tau-1}^{\text{rank}}\times \frac{c_{(k), t+\tau}}{c_{(k), t+\tau-1}}, k=1,2,...,N$ (\eqref{eq: intraday_portfolio_weights_name_dynamics_1})
\State \hspace{\algorithmicindent} \hspace{\algorithmicindent} \hspace{\algorithmicindent} $w_{i, t+\tau}^{\text{name}} \leftarrow w_{i, t+\tau-1}^{\text{name}}\times \frac{c_{i, t+\tau}}{c_{i, t+\tau-1}}, i=1,2,...,N$ (\eqref{eq: intraday_portfolio_weights_name_dynamics_2})
\State \hspace{\algorithmicindent} \hspace{\algorithmicindent} \hspace{\algorithmicindent} \Comment{$t+\tau-1$ and $t+\tau$ are adjacent intraday timestamps}
\State \hspace{\algorithmicindent} \hspace{\algorithmicindent} \hspace{\algorithmicindent} \textbf{if} $\tau\%\mathcal{T}==0$ \textbf{or} end of the trading day \Comment{rebalancing point}
\State \hspace{\algorithmicindent} \hspace{\algorithmicindent} \hspace{\algorithmicindent} \hspace{\algorithmicindent} Calculate $\text{cost}(t+\tau, w_{(k),t}^{\text{rank}})$ by \eqref{eq: cost_intraday_rebalance}
\State \hspace{\algorithmicindent} \hspace{\algorithmicindent} \hspace{\algorithmicindent} \hspace{\algorithmicindent} $V_t \leftarrow V_t - \text{cost}(t+\tau, w_{(k),t}^{\text{rank}})$
\State \hspace{\algorithmicindent} \hspace{\algorithmicindent} \hspace{\algorithmicindent} \hspace{\algorithmicindent} $w_{\mathcal{I}_{(k), t+\tau}, t+\tau}^{\text{name}} \leftarrow w_{(k), t+\tau}^{\text{rank}}, \quad k=1,2,..., N$
\State \hspace{\algorithmicindent} \hspace{\algorithmicindent} \hspace{\algorithmicindent} $\tau\leftarrow \tau + 1$
\State \hspace{\algorithmicindent} \hspace{\algorithmicindent} $V_{t+1} \leftarrow (1+r_{f,t+1})V_t + \sum_{i}w_{i, t+\tau}^{\text{name}}$
\State \hspace{\algorithmicindent} \hspace{\algorithmicindent} $w^{\text{prev}}\leftarrow w_{t+\tau}^{\text{name}}$
\State \hspace{\algorithmicindent} \hspace{\algorithmicindent} $t \leftarrow t+1$
\State \hspace{\algorithmicindent} \textbf{return} $V_t, \quad t=1,2,..., T+1$\\
\State \tcp{Input:}
\State \hspace{\algorithmicindent} \tcp{$w_t^R$: the $l_1$-normalized portfolio weights from either parametric}
\State \hspace{\algorithmicindent} \hspace{\algorithmicindent} \space \space \space \tcp{model (Algorithm \ref{alg: portfolio_weights_parametric_model}) or neural networks (Algorithm \ref{alg: portfolio_weights_neural_network}).}
\tcp{$r_{f, t}$: risk-free rate during the trading day $t$.}
\tcp{$\mathcal{T}$: rebalance interval.}
\tcp{$c_{t+\tau}$: the capitalization processes in name space and rank space at}
\State \hspace{\algorithmicindent} \hspace{\algorithmicindent} \hspace{\algorithmicindent} \tcp{1-minute resolution throughout the trading day $t$.}
\State \hspace{\algorithmicindent} \hspace{\algorithmicindent} \hspace{\algorithmicindent} \tcp{$t$ is the time tick at daily level.}
\State \hspace{\algorithmicindent} \hspace{\algorithmicindent} \hspace{\algorithmicindent} \tcp{$\tau$ is the time tick at minute level.}
\State \tcp{Output:}
\State \hspace{\algorithmicindent} \tcp{$V_t$: the value process of the portfolio (PnL) with weights $w_t^R$.}
\State \tcp{Note:}
\State \hspace{\algorithmicindent} \tcp{The algorithm realizes the formulation in section 2.3.1.}
\State \hspace{\algorithmicindent} \tcp{In essence, it converts portfolio weights from rank to name}
\State \hspace{\algorithmicindent} \hspace{\algorithmicindent} \tcp{at $\mathcal{T}$ minutes interval.}
\end{algorithmic}
\end{algorithm}
\begin{algorithm}[H]
\caption{portfolio metric [\autoref{fig: schematic_formulation_framework}, panel (g)]}\label{alg: portfolio_metric}
\begin{algorithmic}
\State \textbf{Input}: $V_t,\quad t=1,2,..., N$
\State \textbf{Output}: $r_{\text{annual}}$, $\sigma_{\text{annual}}$, $\text{SR}_{\text{annual}}$\\
\State \textbf{Function} portfolio\_metric($V_t, r_f$)
\State \hspace{\algorithmicindent} \textbf{For} year \textbf{in} years\_in\_backtesting:
\State \hspace{\algorithmicindent} \hspace{\algorithmicindent} Locate all $t_1\leq t_2\leq ...\leq t_N$ in year
\State \hspace{\algorithmicindent} \hspace{\algorithmicindent} $r_{t_i} = V_{t_i}/V_{t_{i-1}}-1, \quad i=1,2,..., N$
\State \hspace{\algorithmicindent} \hspace{\algorithmicindent} $r_{\text{annual}}\leftarrow (\prod_{i=1}^N (1+r_{t_i}))^{252/N}-1$
\State \hspace{\algorithmicindent} \hspace{\algorithmicindent} $\sigma_{\text{annual}}\leftarrow \sqrt{252}\times \text{std}(\{r_{t_i}\}_{i=1}^N)$\Comment{std is the standard deviation}
\State \hspace{\algorithmicindent} \hspace{\algorithmicindent} $\text{SR}_{\text{annual}}\leftarrow (r_{\text{annual}}-r_{f, \text{annual}})/\sigma_{\text{annual}}$ \Comment{$r_{f, \text{annual}}$ is annualized risk-free rate}
\State \hspace{\algorithmicindent} \textbf{return} $r_{\text{annual}}, \sigma_{\text{annual}}, \text{SR}_{\text{annual}}$ for all backtesting years\\
\State \tcp{Input:\\$V_t$: the value process (PnL) of the portfolio with weights $w_t^R$.}
\State \hspace{\algorithmicindent} \tcp{$r_{f,t}$: the risk-free rate at the end of trading day $t$.}
\State \tcp{Output:\\The algorithm realizes the formulation in section 2.4.\\$r_{\text{annual}}$: the annualized return for all backtesting years.\\$\sigma_{\text{annual}}$: the annualized volatility for all backtesting years.\\$\text{SR}_{\text{annual}}$: the Sharpe ratio for all backtesting years.}
\end{algorithmic}
\end{algorithm}

\begin{algorithm}[H]
\caption{(Integrated) Statistical arbitrage in name space}\label{alg: stats_arb_name_space}
\begin{algorithmic}
\State \textbf{Input}: $r_t, r_f, K$
\State \textbf{Called algorithm}: Algorithm \ref{alg: market_decomposition}, Algorithm \ref{alg: portfolio_weights_parametric_model}, Algorithm \ref{alg: portfolio_weights_neural_network}, Algorithm \ref{alg: portfolio_metric}
\State \textbf{Output}: $w_t^{R}, V_t$, $\text{SR}_{\text{annual}}$
\State \textbf{Function} statistical\_arbitrage\_in\_name\_space($r_t, r_{f,t}, K)$
\State \hspace{\algorithmicindent} $\epsilon_t, \Phi_t \leftarrow \text{market\_decomposition}(r_t, r_{f,t}, K)$ from Algorithm \ref{alg: market_decomposition}
\State \hspace{\algorithmicindent} \textbf{if} parametric model
\State \hspace{\algorithmicindent} \hspace{\algorithmicindent} $w_t^{R|\text{OU}} \leftarrow \text{portfolio\_weights\_by\_parametric\_model}(\epsilon_t, \Phi_t)$ from Algorithm \ref{alg: portfolio_weights_parametric_model};
\State \hspace{\algorithmicindent} \textbf{if} neural networks
\State \hspace{\algorithmicindent} \hspace{\algorithmicindent} $w_t^{R|\text{NN}} \leftarrow \text{portfolio\_weights\_by\_neural\_network}(\epsilon_t, \Phi_t)$ from Algorithm \ref{alg: portfolio_weights_neural_network}
\State \hspace{\algorithmicindent} Calculate PnL $V_t$ by \eqref{eq: PnL_name_space}
\State \hspace{\algorithmicindent} $r_{\text{annual}}$, $\sigma_{\text{annual}}$, $\text{SR}_{\text{annual}}\leftarrow $ portfolio\_metric($V_t, r_{f,t}$) from Algorithm \ref{alg: portfolio_metric}
\State \hspace{\algorithmicindent} \textbf{return} $w_t^{R}, V_t$, $\text{SR}_{\text{annual}}$\\
\State \tcp{Input:\\$r_t$: dividend-adjusted daily return in name space.}
\State \hspace{\algorithmicindent} \tcp{$r_{f,t}$: risk-free rate at the end of trading day $t$.}
\State \hspace{\algorithmicindent} \tcp{$K$: number of market factors, predetermined by analyzing eigenvalue}
\State \hspace{\algorithmicindent} \hspace{\algorithmicindent} \space \space \tcp{spectrum of the correlation matrix.}
\State \tcp{Output:\\$w_t^R$: the $l_1$-normalized portfolio weights on stock.\\
$V_t$: the value process (PnL) of the portfolio with weights $w_t^R$.\\
$r_{\text{annual}}$, $\sigma_{\text{annual}}$, $\text{SR}_{\text{annual}}$: annualized return, volatility, and Sharpe}
\State \hspace{\algorithmicindent} \hspace{5.25cm}\tcp{ratio.}
\end{algorithmic}
\end{algorithm}

\begin{algorithm}[H]
\caption{(Integrated) Statistical arbitrage in rank space}\label{alg: stats_arb_rank_space}
\begin{algorithmic}
\State \textbf{Input}: $c_t, c_{t+\tau}, r_f, K$
\State \textbf{Called algorithm}: Algorithm \ref{alg: market_decomposition}, Algorithm \ref{alg: portfolio_weights_parametric_model}, Algorithm \ref{alg: portfolio_weights_neural_network}, Algorithm \ref{alg: intraday_rebalance}, Algorithm \ref{alg: portfolio_metric}
\State \textbf{Output}: $w_t^{R}, V_t$, $r_{\text{annual}}$, $\sigma_{\text{annual}}$, $\text{SR}_{\text{annual}}$\\
\State \textbf{Function} statistical\_arbitrage\_in\_name\_space($c_t, r_{f,t}, K)$
\State \hspace{\algorithmicindent} Calculate $\Tilde{r}_t$ by \eqref{eq: def_return_rank}
\State \hspace{\algorithmicindent} $\Tilde{\epsilon}_t, \Tilde{\Phi}_t \leftarrow \text{market\_decomposition}(\Tilde{r}_t, r_{f,t}, K)$ from Algorithm \ref{alg: market_decomposition}
\State \hspace{\algorithmicindent} \textbf{if} parametric model
\State \hspace{\algorithmicindent} \hspace{\algorithmicindent} $w_t^{R|\text{OU}} \leftarrow \text{portfolio\_weights\_by\_parametric\_model}(\epsilon_t, \Phi_t)$ from Algorithm \ref{alg: portfolio_weights_parametric_model};
\State \hspace{\algorithmicindent} \textbf{if} neural networks
\State \hspace{\algorithmicindent} \hspace{\algorithmicindent} $w_t^{R|\text{NN}} \leftarrow \text{portfolio\_weights\_by\_neural\_network}(\epsilon_t, \Phi_t)$ from Algorithm \ref{alg: portfolio_weights_neural_network}
\State \hspace{\algorithmicindent} $V_t\leftarrow$ intraday\_rebalancing($w_t^{R|\text{NN}}, r_{f,t}, \mathcal{T}, c_{t+\tau}$) from Algorithm \ref{alg: intraday_rebalance}
\State \hspace{\algorithmicindent} $r_{\text{annual}}$, $\sigma_{\text{annual}}$, $\text{SR}_{\text{annual}}\leftarrow $ portfolio\_metric($V_t, r_{f,t}$) from Algorithm \ref{alg: portfolio_metric}
\State \hspace{\algorithmicindent} \textbf{return} $w_t^{R}, V_t$, $r_{\text{annual}}$, $\sigma_{\text{annual}}$, $\text{SR}_{\text{annual}}$\\
\State \tcp{Input:\\$c_t$: capitalizations at the end of trading day $t$.\\
$c_{t+\tau}$: capitalization process at 1-minute resolution throughout the} 
\State \hspace{\algorithmicindent} \hspace{\algorithmicindent} \hspace{\algorithmicindent} \tcp{trading day $t$. $t$ is the time tick at daily level.}
\State \hspace{\algorithmicindent} \hspace{\algorithmicindent} \hspace{\algorithmicindent} \tcp{$\tau$ is the time tick at intraday level.}
\State \hspace{\algorithmicindent} \tcp{$r_{f,t}$: risk-free rate at the end of trading day $t$.}
\State \hspace{\algorithmicindent} \tcp{$K$: number of market factors, predetermined by analyzing eigenvalue}
\State \hspace{\algorithmicindent} \hspace{\algorithmicindent} \space \space \tcp{spectrum of the correlation matrix.}
\State \tcp{Output:}
\State \hspace{\algorithmicindent} \tcp{$w_t^R$: the $l_1$-normalized portfolio weights on artificial financial}
\State \hspace{\algorithmicindent} \hspace{\algorithmicindent} \space \space \tcp{instruments that realize $\Tilde{r}_t$.}
\State \hspace{\algorithmicindent} \tcp{$V_t$: the value process (PnL) of the portfolio with weights $w_t^R$.}
\State \hspace{\algorithmicindent} \tcp{$r_{\text{annual}}$, $\sigma_{\text{annual}}$, $\text{SR}_{\text{annual}}$: annualized return, volatility, and Sharpe}
\State \hspace{\algorithmicindent} \hspace{5.25cm}\tcp{ratio.}
\end{algorithmic}
\end{algorithm}

\newpage

\section{Hybrid-Atlas model}\label{sec: hybrid-atlas model}
In this section, we introduce a hybrid-Atlas model that motivates the definition of return in \eqref{eq: def_return_rank} and hints at the qualitative difference between name space and rank space\cite{banner2005atlas, banner2008local, ichiba2011hybrid}.

We study an equity market that consists of $n$ stocks with capitalizations $C(t) = (C_1(t), \ldots, C_n(t))$, where $C_i(t)$ represents the capitalization at time $t$ of the asset with name $i$. We assume that the log-capitalizations $Y_i(t) := \log C_i(t)$, $i = 1, \ldots, n$, satisfy the system of stochastic differential equations:
\begin{equation}
\begin{aligned}
dY_i(t) &= \left( g_{\mathcal{R}_{i, t}} + \gamma_i + \gamma \right) dt + \sum_{j=1}^n \rho_{i,j} \, dW_j(t) \\
&\quad +  \sigma_{\mathcal{R}_{i, t}} Y_i(t) \, dW_i(t), \quad Y_i(0) = y_i, \quad 0 \leq t < \infty
\end{aligned}
\end{equation}
with given initial condition $y = (y_1, \ldots, y_n)'$. We assume that $\gamma=0$ and the system satisfies the stability condition
\begin{equation}
\sum_{k=1}^{n} g_k + \sum_{i=1}^{n} \gamma_i = 0.
\end{equation}

Define the log-capitalization process in rank space, $Z_k(t) = Y_{\mathcal{I}_{k, t}}(t)$. Therefore, the rank return defined in \eqref{eq: def_return_rank} can be viewed as the discrete counterpart of $d Z_k(t)$.

Theorem \ref{theorem: hybrid_atlas_dynamics} highlights the qualitative difference between residual rank space and name space beyond linear transformation.
\begin{lemma}\label{lemma: local_time_summation}
For continuous semimartingales $Y_1, Y_2, ..., Y_n$, the rank process $Z_1, Z_2, ..., Z_n$ are continuous semimartingales, and we have
\begin{equation}
    \sum_{k=1}^n L_t(Z_k) = \sum_{i=1}^n L_t(Y_i), \quad \forall t>0,
\end{equation}
where $L_t(Y) = \frac{1}{2\epsilon}\int_0^t \boldsymbol{1}_{\{-\epsilon<Y_s<\epsilon\}} dY_s$ is the local time accumulated at origin.
\end{lemma}
\begin{proof}
The proof follows \cite{banner2008local} theorem 2.2. 
\end{proof}

\begin{lemma}\label{lemma: hybrid_atlas_dynamics}
For continuous semimartingales $Y_1, Y_2, ..., Y_n$ and their rank process $Z_1, Z_2, ..., Z_n$, we have
\begin{equation}
\begin{aligned}
dZ_k(t) &= \sum_{i=1}^{n} (N_k(t))^{-1} \mathbf{1}_{\{Z_{k}(t) = X_i(t)\}} \, dX_i(t) \\
&\quad + \sum_{j=k+1}^{n} (N_k(t))^{-1} \, dL_t(Z_{k} - Z_{j}) - \sum_{j=1}^{k-1} (N_k(t))^{-1} \, dL_t(Z_{j} - Z_{k}).
\end{aligned}
\end{equation}
, where $S_t(k):= \{i: Y_i(t) = Z_k(t)\}$ and $N_k(t)$ is the cardinality of $S_t(k)$.
\end{lemma}
\begin{proof}
Our proof follows \cite{banner2008local} theorem 2.3 closely.

Define
\begin{equation}
U = \{ u(\cdot) : \mathbb{R}^+ \times \{1, \ldots, n\} \to \{1, \ldots, n\}, Z_k(t) = Y_{u_t(k)}(t), \forall t > 0, \, k = 1, \ldots, n \}.
\end{equation}

For any $\mathbf{j} \in J$, we define $u^{\mathbf{j}}$ of $U$ as
\begin{equation}
u_t^{\mathbf{j}}(k) := \text{the } j_{N_t(k)}\text{-th smallest element of } S_t(k).
\end{equation}
In other words, suppose that at time $t$ precisely $m$ of the processes $X_1, \ldots, X_n$ have rank $k$, denoted as $X_{i_1}, \ldots, X_{i_m}$. Then, $u_t^{\mathbf{j}}(k)$ is the $j_m$-th smallest among the indices $i_1, \ldots, i_m$.

For $u\in U$, we have
\begin{equation}
\begin{aligned}
& Z_k(t) - Z_k(0) = \sum_{i=1}^{n} \int_{0}^{t} \mathbf{1}_{\{u_s(k) = i\}} \, dY_i(s) + \sum_{i=1}^{n} \int_{0}^{t} \mathbf{1}_{\{u_s(k) = i\}} \, d\left( Z_k(s) - Y_i(s) \right)\\
& = \frac{1}{n!} \sum_{i=1}^{n} \int_{0}^{t} \sum_{j \in J} \mathbf{1}_{\{u_s^{\mathbf{j}}(k) = i\}} \, dY_i(s) + \frac{1}{n!} \sum_{i=1}^{n} \int_{0}^{t} \sum_{j \in J} \mathbf{1}_{\{u_s^{\mathbf{j}}(k) = i\}} \, d\left( X_{(k)}(s) - X_i(s) \right).\\
& = \sum_{i=1}^{n} \int_{0}^{t} (N_s(k))^{-1} \mathbf{1}_{\{Z_k(s) = Y_i(s)\}} \, dY_i(s) + \sum_{i=1}^{n} \int_{0}^{t} (N_s(k))^{-1} \mathbf{1}_{\{Z_k(s) = Y_i(s)\}} \, d\left( Z_k(s) - Y_i(s) \right)\\
&= \sum_{i=1}^{n} \int_{0}^{t} (N_s(k))^{-1} \mathbf{1}_{\{Z_k(s) = Y_i(s)\}} \, dY_i(s) + \sum_{i=1}^{n} \int_{0}^{t} (N_s(k))^{-1} \, dL_s\left( (Z_k - Y_i)^+ \right) \\
& \quad \quad - \sum_{i=1}^{n} \int_{0}^{t} (N_s(k))^{-1} \, dL_s\left( (Z_k - Y_i)^- \right).
\end{aligned}
\end{equation}
, where $(\cdot)^+ = \max (\cdot, 0)$ and $(\cdot)^- = \min (\cdot, 0)$. In the second equality, we replace $u$ by $u^j$ for $j\in J$, then summing over all $j$. In the third equality, we use $\sum_{j \in J} \mathbf{1}_{\{u_s^{\mathbf{j}}(k) = i\}} = \frac{n!}{N_s(k)} \, \mathbf{1}_{\{X_{(k)}(s) = X_i(s)\}}$. In the fourth equality, we use lemma \ref{lemma: local_time_summation} and 
\begin{equation}
\begin{aligned}
& \sum_{i=1}^{n} \int_{0}^{t} (N_s(k))^{-1} \mathbf{1}_{\{Z_k(s) = Y_i(s)\}} \, d\left( Z_k(s) - Y_i(s) \right)\\
&= \sum_{i=1}^{n} \int_{0}^{t} (N_s(k))^{-1} \mathbf{1}_{\{Z_k(s) = Y_i(s)\}} \, d\left( (Z_k(s) - Y_i(s))^+ \right) \\
&\quad - \sum_{i=1}^{n} \int_{0}^{t} (N_s(k))^{-1} \mathbf{1}_{\{Z_k(s) = Y_i(s)\}} \, d\left( (Z_k(s) - Y_i(s))^- \right) \\
&= \sum_{i=1}^{n} \int_{0}^{t} (N_s(k))^{-1} \, dL_s\left( (Z_k - Y_i)^+ \right)
 - \sum_{i=1}^{n} \int_{0}^{t} (N_s(k))^{-1} \, dL_s\left( (Z_k - Y_i)^- \right).
\end{aligned}
\end{equation}
\end{proof}

\begin{theorem}\label{theorem: hybrid_atlas_dynamics}
The log-capitalization process in rank space $Z_k$ satisfies
\begin{equation}
\begin{aligned}
dZ_k(t) &= \left( g_{\mathcal{R}_{i, t}} + \gamma_i + \gamma \right) dt + \sum_{j=1}^n \rho_{i,j} \, dW_j(t) + \sigma_{\mathcal{R}_{i, t}} Y_i(t) \, dW_i(t), dY_{\mathcal{I}_{k, t}}(t) \\
& + \frac{1}{2}(d\Lambda^{k, k+1}(t) - d\Lambda^{k-1, k}(t))
\end{aligned}
\end{equation}
, where $\Lambda^{k, k+1}$ is the local time accumulated at the origin by the non-negative semimartingale $Z_k(t)-Z_{k+1}(t)$ defined as
\begin{equation}
    \Lambda^{k, k+1}(t) = \lim_{\varepsilon\downarrow 0} \frac{1}{2\varepsilon} \int_0^t \boldsymbol{1}_{\{-\varepsilon < Z_k(s)-Z_{k+1}(s) < \varepsilon\}} d (Z_k(s) - Z_{k+1}(s))
\end{equation}
\end{theorem}

\begin{proof}
From the lemmas above, we have
\begin{equation}
\begin{aligned}
dZ_k(t) &= \left( g_{\mathcal{R}_{i, t}} + \gamma_i + \gamma \right) dt + \sum_{j=1}^n \rho_{i,j} \, dW_j(t) + \sigma_{\mathcal{R}_{i, t}} Y_i(t) \, dW_i(t), dY_{\mathcal{I}_{k, t}}(t) \\
&\quad + (N_k(t))^{-1} \left[ \sum_{\ell=k+1}^{n} d\Lambda^{k,\ell}(t) - \sum_{\ell=1}^{k-1} d\Lambda^{\ell,k}(t) \right]\\
&= \left( g_{\mathcal{R}_{i, t}} + \gamma_i + \gamma \right) dt + \sum_{j=1}^n \rho_{i,j} \, dW_j(t) + \sigma_{\mathcal{R}_{i, t}} Y_i(t) \, dW_i(t), dY_{\mathcal{I}_{k, t}}(t) \\
& \quad + \frac{1}{2}(d\Lambda^{k, k+1}(t) - d\Lambda^{k-1, k}(t))
\end{aligned}
\end{equation}
, where the second equality follows from \cite{ichiba2011hybrid} lemma 1 that the local times $\Lambda^ {k, l}$ generated by triple or higher-order collisions are identically equal to zero. In other words, $\Lambda^{k, l} = 0$ for $|k-l|\geq 2, 1\leq k, l, \leq n$.
\end{proof}

\newpage
\section{Market neutrality}\label{section: market_neutrality}
The statistical arbitrage portfolios need to satisfy the market neutrality constraint, $w^T\beta = 0$, so that the return of the portfolio is independent of market factors
\begin{equation}
    w^T(r_t - r_f) = w^T(\beta_tF_t + \epsilon_t) = w^T \epsilon_t.
\end{equation}
The following theorem proves that the market neutrality of the portfolio constructed in section \ref{section: market_decomposition} satisfies the constraints.

\begin{theorem}
If the portfolio weights satisfies \autoref{eq: portfolio_weights_conversion_name_space} or \autoref{eq: portfolio_weights_conversion_rank_space}, it is market neutral.
\end{theorem}

\begin{proof}
We denote the return matrix $R_t = (r_{t-T+1}, r_{t-T+2}, ..., r_{t})\in \mathbb{R}^{N\times T}$. Assume singular value decomposition of $R_t - R_f$,
\begin{equation}
    R_t - R_f = U\Sigma V^T
\end{equation}
, where $R_f\in \mathbb{R}^{1\times T}$ is the risk-free rate, $U\in \mathbb{R}^{N\times N}$, $\Sigma\in\mathbb{R}^{N\times T}$, and $V^T\in \mathbb{R}^{T\times T}$. Then, the factors and loadings in \autoref{eq: market_decomposition_name_main} and $\omega_t$ in \autoref{eq: market_decomposition_name_factor} becomes
\begin{equation}
F_t = \begin{pmatrix}v_1^T\\v_2^T\\...\\v_K^T\end{pmatrix}, \quad \beta_t = \begin{pmatrix}u_1, u_2,...,u_K\end{pmatrix}\begin{pmatrix}\sigma_1 & & \\& ... &\\ & & \sigma_K\end{pmatrix}, \quad \omega_t = \begin{pmatrix}\sigma_1^{-1} & & \\& ... &\\ & & \sigma_K^{-1}\end{pmatrix}\begin{pmatrix}u_1^T\\u_2^T\\...\\u_K^T\end{pmatrix}
\end{equation}
, where $u_i$ and $v_i$ are the $i$-th column of matrix $U$ and $V$. Then, due to the orthogonality between $U$ and $V$, 
\begin{equation}
\Phi_t \beta_t = (I-\beta_t\omega_t)\beta_t = \beta_t - \beta_t (\omega_t\beta_t) = \beta_t - \beta_t = 0,
\end{equation}
. Therefore, 
\begin{equation}
    (w_t^{R})^T\beta = (w_t^{\epsilon})^T\Phi_t\beta = 0
\end{equation}
\end{proof}

\newpage
\section{Intraday rebalancing}\label{section: intraday_rebalancing}
The portfolio weights calculated in the rank space are assigned to artificial financial instruments that yield rank returns in continuous-time limits defined in \eqref{eq: def_return_rank}. To make the constructed portfolio practically implementable, it is necessary to convert these portfolio weights into stock-based portfolios in name space.

A naive approach is to assign the portfolio weights with correspondence between ranks and names at the end of each trading day and hold the portfolio throughout the following trading day,
\begin{equation}
    w_{i, t} = \sum_{k=1}^N w_{(k), t}\boldsymbol{1}_{\{\mathcal{R}_{i,t}=k\}}, \quad \quad i=1,2,..., N.
\end{equation}
Unfortunately, this straightforward conversion will not retain the advantages of statistical arbitrage in rank space, because the performance of the derived portfolio will essentially still depend on returns in name space rather than the rank returns in continuous-time limit. As indicated by \eqref{eq: def_return_rank}, the returns in name and rank space start to diverge in the event of rank switching that frequently occurs for most ranks at an intraday frequency. It indicates that an effective conversion strategy must appropriately respond to the rank-switching events.

Consequently, we propose an intraday rebalancing mechanism in section \ref{section: intraday_rebalancing_main_text}. This mechanism performs conversion from rank space to name space at a frequency that matches rank-switching events, even though it results in higher transaction costs due to more frequent trading. In the following, we carry out an in-depth analysis in a two-stock system to emphasize the crucial role of rank switching and the rebalancing interval in determining the cost of conversion.

\subsection{Portfolio rebalancing through rank switching of two stocks}
\begin{figure}[h!]
    \centering
    \includegraphics[scale=0.25]{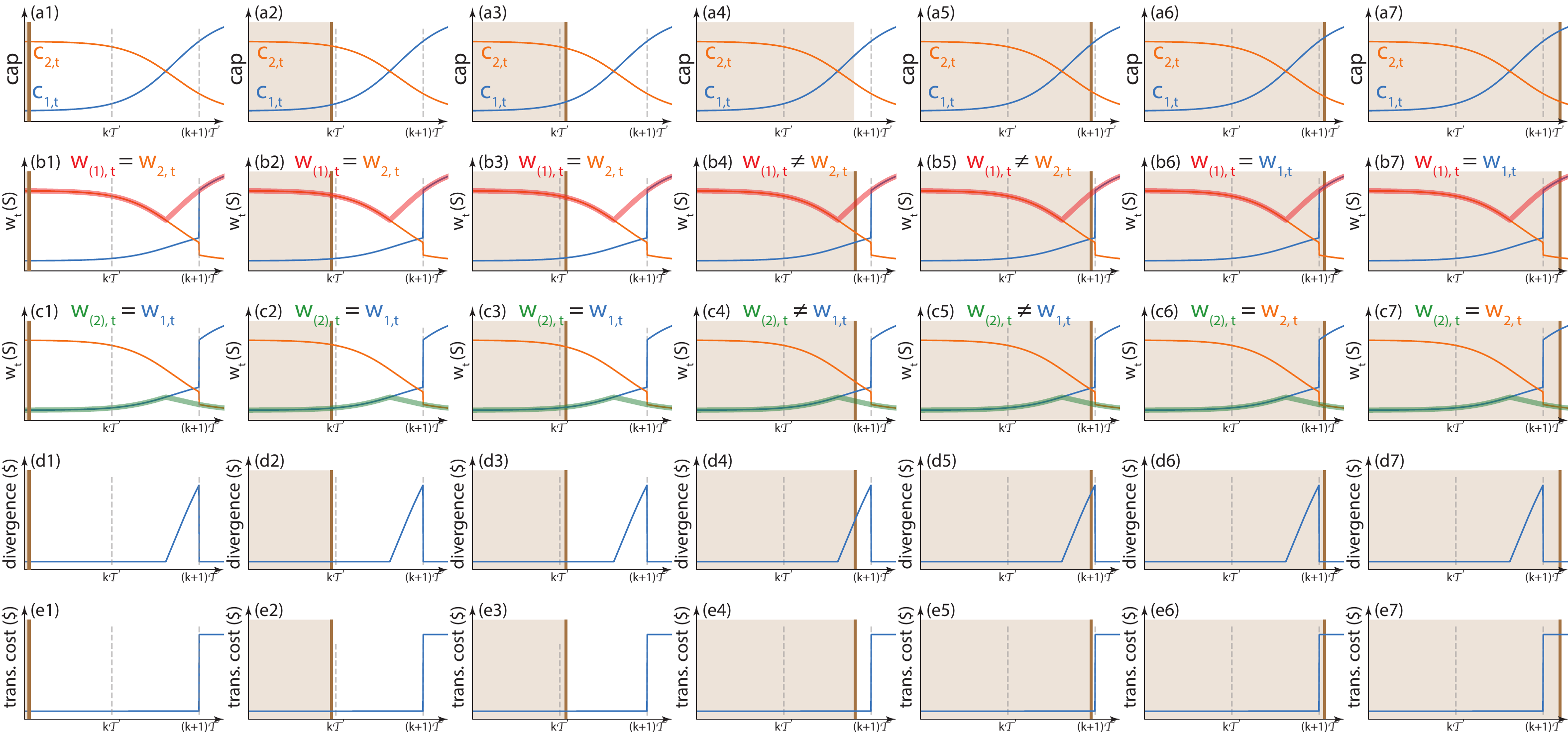}
    \caption{\textbf{Schematic for the intraday rebalancing through rank switching of two stocks.} Here, we examine a two-stock system and highlight the critical role of the rank switching. We consider two capitalization processes $c_{t,1}$ and $c_{t,2}$ that maintain their ranks during the rebalancing interval $((k-1)\mathcal{T}, k\mathcal{T}]$ and switch their ranks during $(k\mathcal{T}, (k+1)\mathcal{T}]$ (panel (a1-a7)). The red lines in panel (b1-b7) and green lines in panel (c1-c7) show the dollar-valued portfolio in rank space $w_{(1),\tau}$ and $w_{(2),\tau}$ respectively, where $w_{(k), t}$ denotes the dollar portfolio weights on the stock that occupies $k$-th rank in capitalization. The orange line in panel (b1-b7) and blue line in panel (c1-c7) show the dollar-valued portfolio in name space $w_{1,\tau}$ and $w_{2,\tau}$ respectively, where $w_{i,\tau}$ denotes the dollar portfolio weights on the stock that had $i$-th name index. We further calculate and present in panel (d1-d7) the divergence between the total dollar portfolio weights in rank space and in name space, defined as $w_{(1),t}+w_{(2),t}-w_{1,t}-w_{2,t}$. Panel (e1-e7) shows the cumulative cost from bid-ask spread arising from the active trading at the rebalancing point.}
    \label{fig: schematic_intraday_rebalance}
\end{figure}

\noindent To elucidate the pivotal role of rank switching in our intraday rebalancing strategy, we examine a two-stock system depicted in \autoref{fig: schematic_intraday_rebalance}, where the two capitalization processes $c_{t,1}$ and $c_{t,2}$ maintain their ranks during the rebalancing interval $((k-1)\mathcal{T}, k\mathcal{T}], k\in\mathbb{N}$ and swap their ranks during $(k\mathcal{T}, (k+1)\mathcal{T}], k\in\mathbb{N}$ (panels (a1-a7)). The red lines in panels (b1-b7) and green lines in panels (c1-c7) show the dollar portfolio weight on the stock that occupies $k$-th rank in capitalization, $w_{(k),\tau}, k=1,2$. The orange lines in panels (b1-b7) and blue lines in panels (c1-c7) show the dollar portfolio weight on the stock that has $i$-th name index, $w_{i,\tau}, i=1,2$. We further calculate and present in panels (d1-d7) the divergence between the total dollar portfolio weights in rank space and the total dollar portfolio weights in name space, defined as $w_{(1),t}+w_{(2),t}-w_{1,t}-w_{2,t}$. Panels (e1-e7) shows the cumulative cost from the bid-ask spread arising from the active trading at the rebalancing point. We highlight several representative timestamps elaborated below.

(i) $t=(k-1)\mathcal{T}^+$ in panels (a1-e1): We invest $w_{(2),t}$ on stock 1 and $w_{(1), t}$ on stock 2 since $c_{1,t}<c_{2,t}$. Therefore, $w_{(1),t}=w_{2,t}$ and $w_{(2),t}=w_{1,t}$;

(ii) $t=k\mathcal{T}$ in panel (a2-e2): The processes evolve towards the rebalancing point $k\mathcal{T}$. The relationship $w_{(1),t}=w_{2,t}, w_{(2),t}=w_{1,t}$ maintains because there is no rank-swapping between stock 1 and stock 2;

(iii) $t=k\mathcal{T}^+$ in panels (a3-e3): At the rebalancing point, no active trading is needed as $w_{(1),t}=w_{2,t}, w_{(2),t}=w_{1,t}$ for $i=1,2$, and therefore no divergence (latency cost) or cost from bid-ask spread incurred;

(iv) $k\mathcal{T} < t \leq (k+1)\mathcal{T}$ in panels (a4-e4, a5-e5): The processes evolve towards the rebalancing point $(k+1)\mathcal{T}$. However, because of the rank switch in capitalization between stock 1 and 2 during the interval, the dollar-valued portfolio for rank and the dollar-valued portfolio for name start diverging, i.e. $w_{(1),t}\neq w_{2,t}, w_{(2),t}\neq w_{1,t}$ and reaches a maximum at the next rebalancing point $(k+1)\mathcal{T}$ (panel (d4, d5));

(v) $t=(k+1)\mathcal{T}^+$ in panels (a6-e6): We carry out active trading to rebalancing the portfolio such that $w_{(1),t}= w_{1,t}, w_{(2),t}= w_{2,t}$. This requires cash reserve to compensate (i) divergence $w_{(1), t}+w_{(2),t}-w_{1,t}-w_{2,t}$ (panel (d6)), and (ii) cost from the bid-ask spread (e6);

(vi) $t>(k+1)\mathcal{T}^+$ in panels (a7-e7): the system continues to evolve with $w_{(1),t}= w_{1,t}, w_{(2),t}= w_{2,t}$, and the divergence becomes zero.

From the detailed analysis above, the need for active trading stems from their rank switching during the balance interval. Furthermore, the latency cost is tied to the divergence of total dollar portfolio weights between rank space and name space, $w_{(1),t}^{\text{rank}}+w_{(2),t}^{\text{rank}}-w_{1,t}^{\text{name}}-w_{2,t}^{\text{name}}$. This divergence increases with the interval between the time for rank switching and the time for the subsequent rebalancing point, suggesting that decreasing rebalancing intervals $\mathcal{T}$ might reduce the risk of large transaction costs by minimizing latency costs.

\begin{figure}[h!]
    \centering
    \includegraphics[scale=0.3]{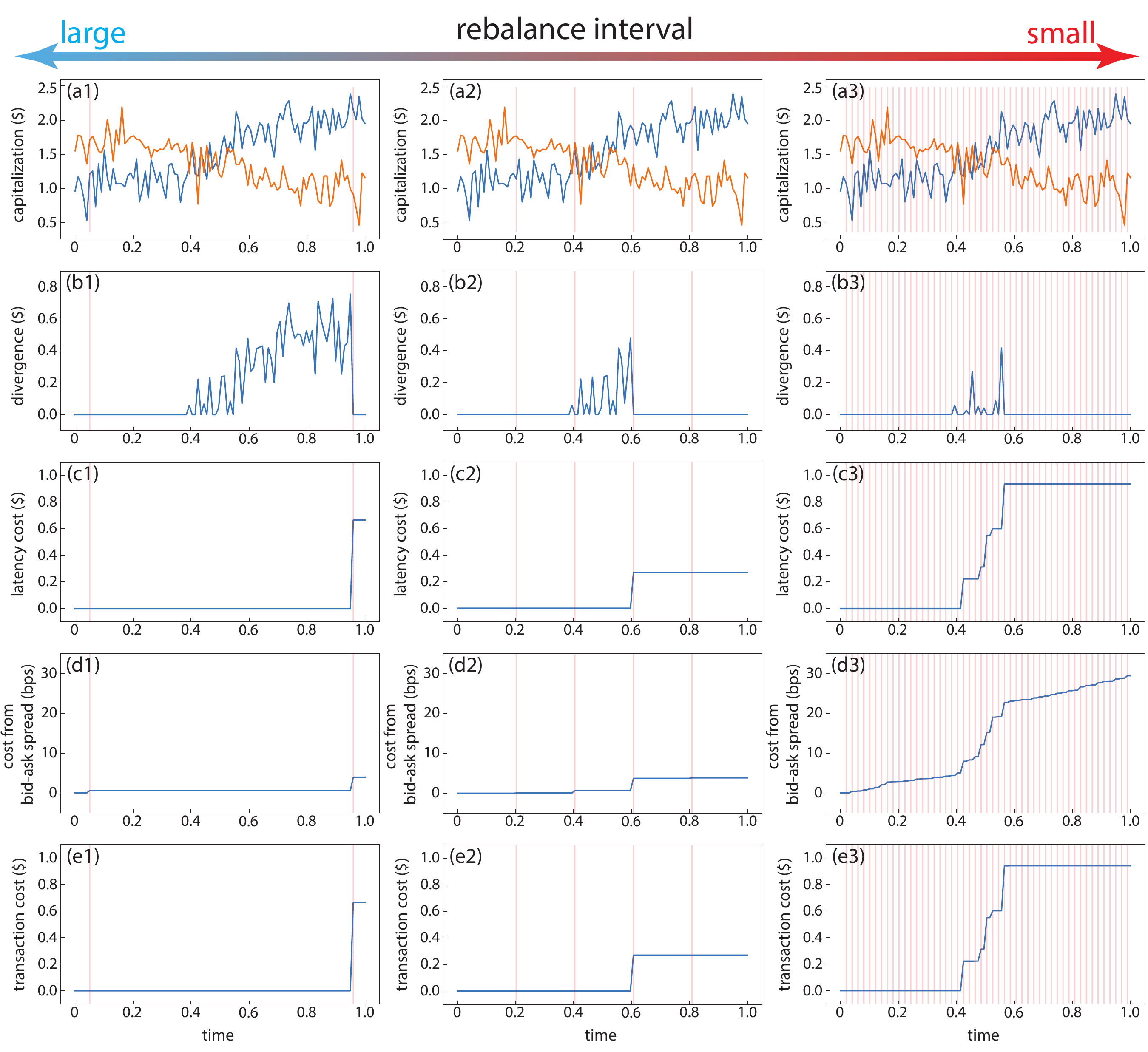}
    \caption{\textbf{Transaction cost dependence on the rebalancing interval.} This figure illustrates how transaction costs incurred by intraday rebalancing are influenced by different rebalancing intervals: large (panel (a1-e1)), medium (panel (a2-e2)), and small (panel (a1-a3)). The red vertical lines mark the rebalancing points across all panels. \textbf{(a1-a3)} The capitalization processes from two stocks. \textbf{(b1-b3)} The divergence between the total dollar portfolio weights in rank space and name space, $w_{(1),t}+w_{(2),t}-w_{1,t}+w_{2,t}$. The dollar portfolio weights in rank space $w_{(1),t}, w_{(2),t}$, and in name space $w_{1,t}, w_{2,t}$ are calculated based on capitalization processes in (a) following the intraday rebalancing strategy similar to \autoref{fig: schematic_intraday_rebalance}. The maximum divergence decreases as rebalancing interval increases, aiding risk control. \textbf{(c1-c3)} The cumulative latency cost required to compensating divergences at the rebalancing points. Each point of rebalancing incurs a latency cost equal to the divergence between the total dollar portfolio weights in rank space and name space. \textbf{(d1-d3)} The cumulative cost from the bid-ask spread due to active trading at each rebalancing point. \textbf{(e1-e3)} The cumulative transaction cost. The cumulative transaction costs are the sum of latency costs and transaction costs. The medium rebalancing interval results in the lowest transaction costs while maintaining a manageable divergence, illustrating the importance of choosing an appropriate rebalancing interval.}

    \label{fig: schematic_intraday_dependence_frequency_dependence}
\end{figure}

However, the situation becomes more complex when considering the fluctuating nature of the capitalization process. In the scenario where two adjacent capitalization processes frequently switch ranks, as shown in \autoref{fig: schematic_intraday_dependence_frequency_dependence}, trading too frequently in response to the instantaneous rank changes can incur substantial, yet unnecessary costs from bid-ask spread. To illustrate this, we present a similar two-particle system where fluctuating capitalization processes cross their paths (\autoref{fig: schematic_intraday_dependence_frequency_dependence}(a1-a3)). We calculate dollar portfolio weights in name space and rank space according to the aforementioned intraday rebalancing strategy, and analyze the divergence of total dollar portfolio weights between rank space and name space (\autoref{fig: schematic_intraday_dependence_frequency_dependence}(b1-b3)), the cumulative latency costs (\autoref{fig: schematic_intraday_dependence_frequency_dependence}(c1-c3)), the cumulative costs from bid-ask spread (\autoref{fig: rebalance_interval_dependence}(d1-d3)), and the cumulative transaction cost (\autoref{fig: schematic_intraday_dependence_frequency_dependence}(e1-e3)). We consider three scenarios under large, medium, and small rebalancing intervals. Remarkably, our findings underscore a return-risk trade-off: frequent trading (short rebalancing interval) yields lower divergence and hence lower risk but incurs higher costs from the bid-ask spread, whereas less frequent trading (large rebalancing interval) results in higher divergence and risk but lower bid-ask costs. Thus, selecting an appropriate rebalancing interval is crucial for minimizing overall transaction costs by balancing between latency costs and costs from bid-ask spread. Indeed, we observe a strong dependence on the profit and loss (PnL) with different intraday rebalance intervals in our empirical analysis (\autoref{fig: rebalance_interval_dependence}).

\clearpage
\section{Benchmark parametric model}\label{sec: ou_process}
\begin{figure}[h!]
    \centering
    \includegraphics[scale=0.7]{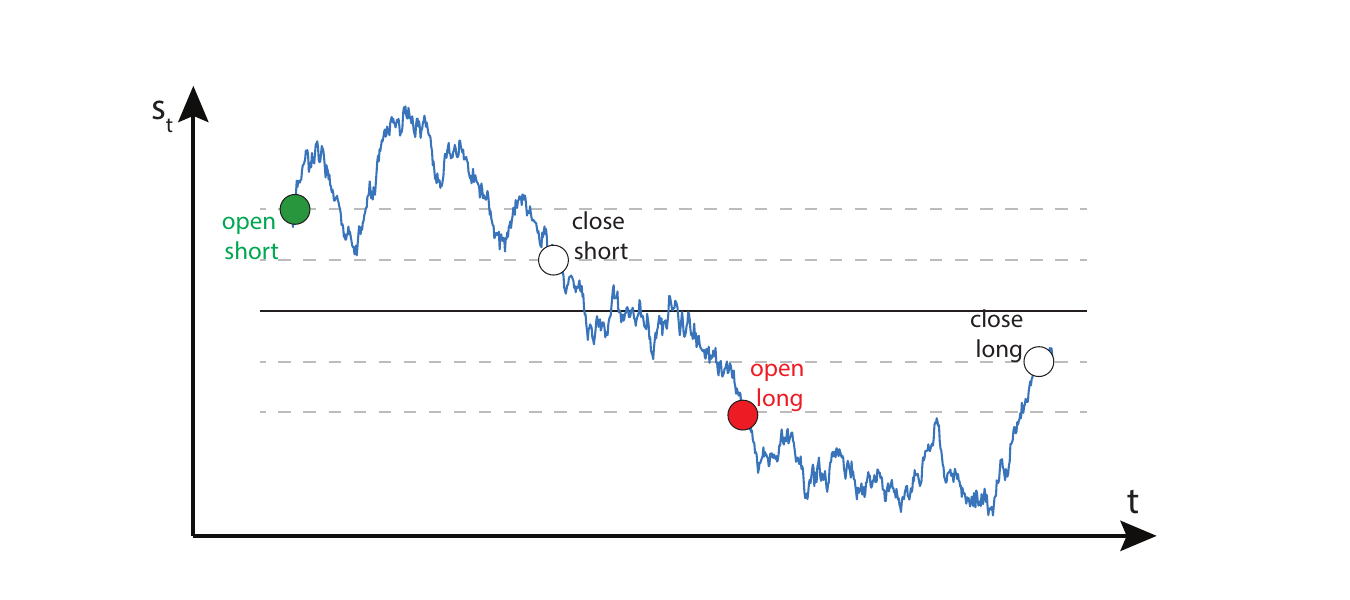}
    \caption{\textbf{Schematic for the parametric model.} The blue line shows the trading signal $s_t$. We open short/long positions when observing large positive/negative signals and close positions when the trading signals mean-revert to zero.}
    \label{fig: schematic_parametric_model}
\end{figure}

Our parametric model serves as the benchmark for DNNs comparison and follows closely to the framework proposed by Avellaneda and Lee\cite{avellaneda2010statistical} and refined by Yeo and Papanicolaou\cite{yeo2017risk}. This model applies to both name space and rank space, depending on whether $x_t^L$ or $\Tilde{x}_t^L$ is chosen as the input. We first fit $x_t^L$ to an OU process $X_t$ governed by the stochastic differential equation 
\begin{equation}\label{eq: def_OU_process}
    dX_t = \frac{1}{\tau} (\mu-X_t) dt + \sigma dB_t,
\end{equation}
where $\tau$ is the mean-reverting time, $\mu$ is the long-term average of $X_t$, $B_t$ is the standard Brownian motion, and $\sigma$ is its volatility. Subsequently, we calculate the trading signal in name space
\begin{equation}\label{eq: trading_signal_OU_process}
    s^{\text{OU}}_{i,t} = \frac{x_{i,t}-\hat{\mu}_i}{\hat{\sigma_i}},
\end{equation}
where $\hat{\mu}_i, \hat{\sigma}_i$ are the maximum likelihood estimator of $\mu$ and $\sigma$\cite{avellaneda2010statistical} and $x_{i,t}$ is the terminal cumulative residual return at time $t$,
\begin{equation} \label{eq: def_terminal_cumulative_residual_return}
x_{i,t} = \sum_{j=1}^L \epsilon_{i, t-L+j}.
\end{equation}
We also include the estimated mean-reverting time $\hat{\tau}$ to effectively filter the trading opportunities\cite{yeo2017risk}. The details of parameter estimation are presented in the Appendix. We open short/long positions when observing large positive/negative signals and close positions when the trading signals mean-revert close to zero (schematic in \autoref{fig: schematic_parametric_model}). Following the principle, the portfolio weights in residual space, $w_t^{\epsilon|\text{OU}, \text{name/rank}}$ become
\begin{equation}\label{eq: portfolio_weights_epsilon_OU_process}
    w_{i,t}^{\epsilon|\text{OU}, \text{name/rank}}= 
    \begin{cases}
    -1, &\text{if}\quad w_{i,t-1}^{\epsilon|\text{OU}}=0, \quad s_{i,t}^{\text{OU}}>c_{\text{thresh-open}}, \quad \hat{\tau_i}<30 \text{ days}\\
    1, & \text{if}\quad w_{i, t-1}^{\epsilon|\text{OU}}=0, \quad s_{i,t}^{\text{OU}}<-c_{\text{thresh-open}},\quad \hat{\tau_i}<30 \text{ days}\\
    1, &\text{if}\quad w_{i,t-1}^{\epsilon|\text{OU}}= 1, \quad s_{i,t}^{\text{OU}} > c_{\text{thresh-close}}, \quad \hat{\tau_i}<30 \text{ days}\\
    -1, &\text{if}\quad w_{i,t-1}^{\epsilon|\text{OU}}= -1, \quad s_{i,t}^{\text{OU}} > c_{\text{thresh-close}}, \quad \hat{\tau_i}<30 \text{ days}\\
    0, & \text{otherwise}
    \end{cases}
\end{equation}
For our back-testing, the parameters are set as follows:
\begin{equation}
    c_{\text{thresh-open}} = 1.25,\quad c_{\text{thresh-close}}=0.5,
\end{equation}
in accordance with \cite{avellaneda2010statistical, yeo2017risk}. After calculating $w_t^{\epsilon|\text{OU}, \text{name/rank}}$, the conversion to portfolio weights in equity space, $w_t^{R|\text{OU}, \text{name/rank}}$, straightforwardly follow the \eqref{eq: portfolio_weights_conversion_name_space} in name space 
\begin{equation}\label{eq: portfolio_weights_equity_space_OU}
w_t^{R|\text{OU}, \text{name}} = \Phi_t^T w_t^{\epsilon|\text{OU}, \text{name}}
\end{equation}
and from \eqref{eq: portfolio_weights_conversion_rank_space} in rank space,
\begin{equation}\label{eq: portfolio_weights_equity_space_OU}
w_t^{R|\text{OU}, \text{rank}} = \Tilde{\Phi}_t^T w_t^{\epsilon|\text{OU}, \text{name}}.
\end{equation}
The practical implementation of the parametric model is summarized in Algorithm \ref{alg: portfolio_weights_parametric_model} along with a schematic in panel (d1, e1) in \autoref{fig: schematic_formulation_framework}.

\clearpage
\section{Non-parametric analysis on mean-reversion of residual returns}\label{sec: non_parametric_analysis_on_mean_reversion}
The superior mean-reverting behavior in rank space is further demonstrated by comparing the empirical distribution of normalized cumulative residual returns, $\Tilde{x}_t^L$ calculated from name space and rank space. We define the normalized cumulative residual returns as follows:
\begin{equation}\label{eq: def_normalized_cumulative_residual_return}
    \hat{x}_t^L = (\hat{x}_{t-L+1}, \hat{x}_{t-L+2}, ..., \hat{x}_t), \quad\text{where }\hat{x}_{t-L+\alpha} = \frac{1}{{\hat{\sigma}_t^L}\sqrt{\alpha}}\sum_{j=1}^\alpha \epsilon_{t-L+j}
\end{equation}
, where $\hat{\sigma}_t^L$ is the estimated standard deviation of $\{\epsilon_{t-L+j}\}_{j=1}^L$. Suppose the residual returns follow uncorrelated, normal distribution, i.e. $\{\epsilon_{t-L+j}\}_{j=1}^L\sim \mathcal{N}(\boldsymbol{0}, \boldsymbol{I}_L)$, the derived cumulative residual returns $x_t^L$ will follow a standard Brownian motion and the normalized cumulative residual return defined in \eqref{eq: def_normalized_cumulative_residual_return} will be normally distributed, i.e. $\hat{x}_{t-L+\alpha}\sim \mathcal{N}(0, 1)$, $\forall \alpha=1, 2, ... L$. Consequently, it will serve as a measure of mean-reversion of the difference in probability density function (p.d.f.) between the empirical observations on market and the normal distribution,
\begin{equation}
    \text{p.d.f.}(\hat{x}_{t-L+\alpha}) - \frac{1}{\sqrt{2\pi}}\exp(-\frac{\hat{x}_{t-L+\alpha}^2}{2}), \quad \alpha=1,2,..., L
\end{equation}

\begin{figure}[h!]
    \centering
    \includegraphics[width=\columnwidth]{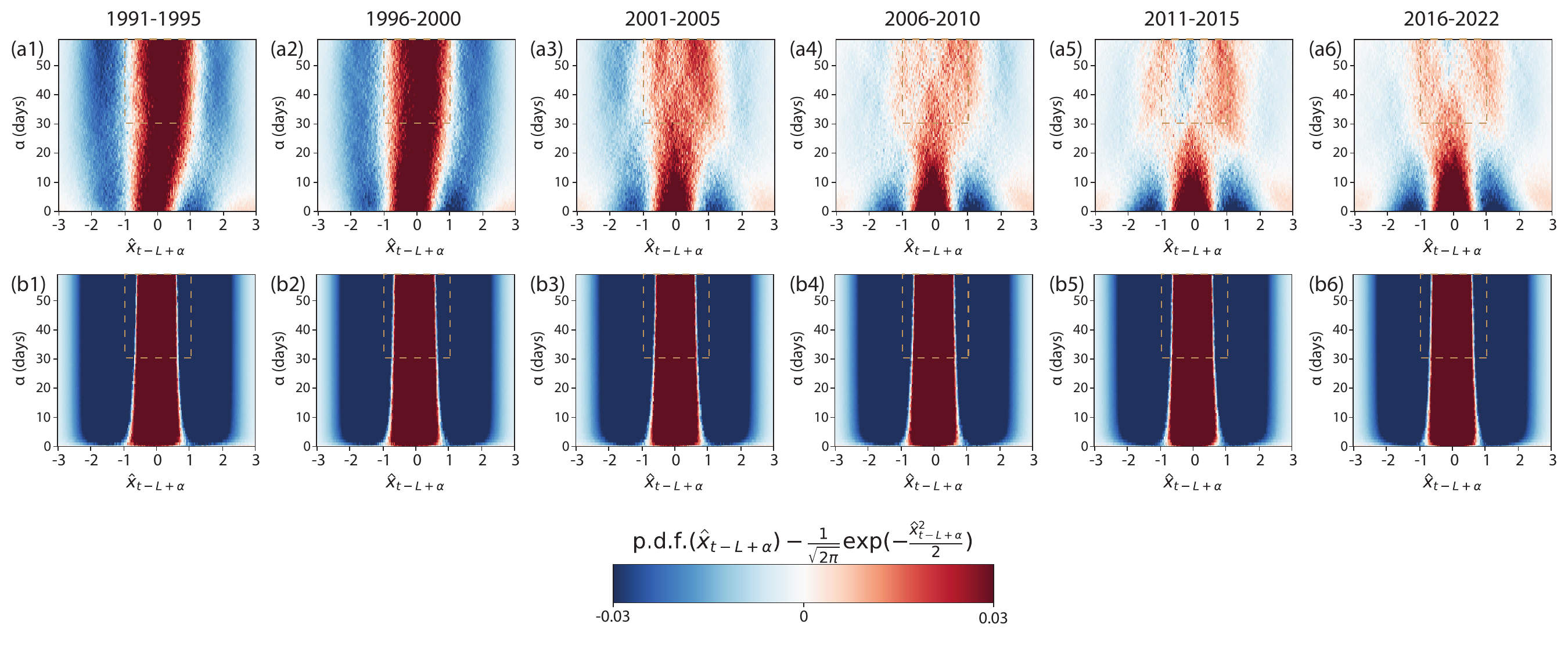}
    \caption{\textbf{Empirical distributions of normalized cumulative residual returns: name space versus rank space.} The cumulative residual returns $x_t^L$ are normalized according to \eqref{eq: def_normalized_cumulative_residual_return}. This normalization facilitates the conversion of comparisons between trajectories of $x_t^L$ and Brownian motion into comparisons of the probability density of the empirical distribution of $\hat{x}_t^L$ against the normal distribution. Consequently, we present the difference between the empirical probability density of $\hat{x}_t^L$ and standard normal distribution. The empirical probability density of $\hat{x}_t^L$ is evaluated across a series of five-year periods from  1991 (left) to 2022 (right) in both \textbf{(a1-a6)} name space and \textbf{(b1-b6)} rank space. The dashed brown box highlights the critical regime where a dominating red color indicates a more concentrated distribution of $x_{t-L+\alpha}$ for large $\alpha$. This concentration signals superior mean-reversion capabilities, particularly evident in rank space throughout the last thirty years. Furthermore, the distribution of $\hat{x}_t^L$ in name space evolves significantly, indicating a progressively deteriorating mean-reversion in name space after 2010. This echoes the relatively poor performance in our backtesting. The stark contrast in $\hat{x}_t^L$ supports the strategic advantage of operating in rank space for statistical arbitrage.}

    \label{fig: epdf_cumulative_residual_return}
\end{figure}

The difference is accessed over a series of five-year periods from 1991 to 2022 for both name space (\autoref{fig: epdf_cumulative_residual_return}(a1-a6)) and rank space (\autoref{fig: epdf_cumulative_residual_return}(b1-b6)). A more concentrated distribution of $\hat{x}_{t-L+\alpha}$ than Brownian motion indicates good mean-reverting behavior, especially for large $\alpha$. This is particularly evident in rank space, where a robust dominance of red color in the heatmaps for large $\alpha$ regime (highlighted in dashed boxes in \autoref{fig: epdf_cumulative_residual_return}) underscores the concentrated nature of $\hat{x}_t^L$ in rank space. Such behavior provides critical evidence of a more robust mean-reversion of residual returns in rank space. In stark contrast, the similar dominance by red color in name space was evident in 1990s (\autoref{fig: epdf_cumulative_residual_return}(a1, a2)), but progressively deteriorated since 2000s (\autoref{fig: epdf_cumulative_residual_return}(a5-a6)), and finally disappears after 2010s (\autoref{fig: epdf_cumulative_residual_return}(a5-a6)). This marks the deterioration of the mean-reversion of $x_t^L$ in name space after the 2010s, echoing the failure of profiting from conventional statistical arbitrage strategies after 2010s.

In summary, our non-parametric analysis here demonstrates that the rank space exhibits more robust mean-reverting behavior compared to name space, echoing the parametric analysis on mean-reverting time in the main text.

\clearpage
\section{Portfolio performance}\label{sec: portfolio_performance}
\subsection{Annualized return, volatility, and Sharpe ratio}\label{sec: portfolio_performance_sharp}
We calculate the annualized return, volatility, and Sharpe ratio derived from the PnL $V_t$ in \autoref{fig: portfolio_performance_summary}. The annualized summary statistics without 
and with transaction costs are presented in \autoref{table: portfolio_performance_wo_tc} and \autoref{table: portfolio_performance_with_tc}, respectively, where we consider corresponding portfolio weights $w_t^R$ calculated by four scenarios: (i) the parametric benchmark model in name space in panel (a), (ii) the parametric benchmark model in rank space in panels (b,c), (iii) DNNs in name space in panel (d), and (iv) DNNs in rank space in panels (e, f). 

\begin{table}[h!]
\scriptsize
\begin{tabular}{|c|lll|lll|lll|lll|}
\hline
\multirow{2}{*}{\textbf{year}} & \multicolumn{3}{c|}{\textbf{\begin{tabular}[c]{@{}c@{}}Name space\\ parametric model\end{tabular}}}       & \multicolumn{3}{c|}{\textbf{\begin{tabular}[c]{@{}c@{}}rank space \\ parametric model\end{tabular}}}      & \multicolumn{3}{c|}{\textbf{\begin{tabular}[c]{@{}c@{}}name space \\ neural networks\end{tabular}}}       & \multicolumn{3}{c|}{\textbf{\begin{tabular}[c]{@{}c@{}}rank space \\ neural networks\end{tabular}}}       \\ \cline{2-13} 
                               & \multicolumn{1}{c}{\textbf{return}} & \multicolumn{1}{c}{\textbf{vol}} & \multicolumn{1}{c|}{\textbf{SR}} & \multicolumn{1}{c}{\textbf{return}} & \multicolumn{1}{c}{\textbf{vol}} & \multicolumn{1}{c|}{\textbf{SR}} & \multicolumn{1}{c}{\textbf{return}} & \multicolumn{1}{c}{\textbf{vol}} & \multicolumn{1}{c|}{\textbf{SR}} & \multicolumn{1}{c}{\textbf{return}} & \multicolumn{1}{c}{\textbf{vol}} & \multicolumn{1}{c|}{\textbf{SR}} \\ \hline
\textbf{2007}                  & 2.87\%                              & 0.02                             & 1.38                             & 8.52\%                              & 0.04                             & 1.90                             & -8.37\%                             & 0.10                             & -0.87                            & 79.41\%                             & 0.14                             & 5.62                             \\
\textbf{2008}                  & 7.42\%                              & 0.04                             & 1.70                             & 25.44\%                             & 0.06                             & 3.96                             & -6.98\%                             & 0.13                             & -0.52                            & 239.88\%                            & 0.38                             & 6.36                             \\
\textbf{2009}                  & 7.01\%                              & 0.03                             & 2.02                             & 27.81\%                             & 0.08                             & 3.57                             & 3.91\%                              & 0.13                             & 0.29                             & 414.09\%                            & 0.35                             & 11.72                            \\
\textbf{2010}                  & -0.49\%                             & 0.02                             & -0.24                            & 31.73\%                             & 0.05                             & 6.21                             & -0.02\%                             & 0.08                             & 0.00                             & 222.37\%                            & 0.19                             & 11.41                            \\
\textbf{2011}                  & 1.91\%                              & 0.02                             & 0.85                             & 40.14\%                             & 0.06                             & 7.12                             & 12.84\%                             & 0.08                             & 1.67                             & 126.84\%                            & 0.22                             & 5.76                             \\
\textbf{2012}                  & -0.40\%                             & 0.02                             & -0.22                            & 41.06\%                             & 0.05                             & 8.20                             & 7.35\%                              & 0.07                             & 1.00                             & 162.37\%                            & 0.20                             & 8.20                             \\
\textbf{2013}                  & 1.19\%                              & 0.02                             & 0.70                             & 27.92\%                             & 0.05                             & 5.74                             & 8.34\%                              & 0.07                             & 1.14                             & 289.73\%                            & 0.21                             & 13.96                            \\
\textbf{2014}                  & 3.65\%                              & 0.02                             & 2.07                             & 43.82\%                             & 0.05                             & 8.84                             & -3.24\%                             & 0.07                             & -0.49                            & 168.89\%                            & 0.14                             & 12.07                            \\
\textbf{2015}                  & 0.81\%                              & 0.02                             & 0.41                             & 41.78\%                             & 0.06                             & 7.44                             & 0.71\%                              & 0.08                             & 0.09                             & 137.95\%                            & 0.19                             & 7.10                             \\
\textbf{2016}                  & 2.79\%                              & 0.02                             & 1.43                             & 61.86\%                             & 0.07                             & 9.51                             & 8.58\%                              & 0.10                             & 0.90                             & 293.85\%                            & 0.27                             & 11.02                            \\
\textbf{2017}                  & 1.56\%                              & 0.02                             & 0.93                             & 30.58\%                             & 0.04                             & 7.04                             & 9.88\%                              & 0.07                             & 1.35                             & 208.30\%                            & 0.17                             & 12.10                            \\
\textbf{2018}                  & 3.07\%                              & 0.02                             & 1.44                             & 27.78\%                             & 0.05                             & 5.71                             & 3.67\%                              & 0.06                             & 0.60                             & 151.91\%                            & 0.19                             & 7.83                             \\
\textbf{2019}                  & 4.50\%                              & 0.02                             & 2.44                             & 41.42\%                             & 0.05                             & 8.39                             & -6.84\%                             & 0.06                             & -1.07                            & 175.91\%                            & 0.20                             & 8.59                             \\
\textbf{2020}                  & 1.56\%                              & 0.04                             & 0.39                             & 25.06\%                             & 0.09                             & 2.89                             & 1.24\%                              & 0.09                             & 0.14                             & 307.81\%                            & 0.28                             & 10.84                            \\
\textbf{2021}                  & -0.67\%                             & 0.02                             & -0.28                            & 37.60\%                             & 0.06                             & 6.21                             & -2.32\%                             & 0.08                             & -0.30                            & 177.60\%                            & 0.25                             & 7.11                             \\
\textbf{2022}                  & 1.05\%                              & 0.03                             & 0.36                             & 36.79\%                             & 0.07                             & 5.53                             & 28.84\%                             & 0.09                             & 3.29                             & 146.86\%                            & 0.29                             & 5.00                             \\ \hline
\textbf{Avg}                   & 2.36\%                              & 0.02                             & 0.96                             & 34.33\%                             & 0.06                             & 6.14                             & 3.60\%                              & 0.08                             & 0.45                             & 206.49\%                            & 0.23                             & 9.04                             \\ \hline
\end{tabular}
\caption{\textbf{Portfolio performance without transaction costs.} The portfolios in rank space consistently outperform their counterparts in name space, both with the parametric model and neural networks. The neural networks improve the portfolio performance in rank space dramatically, in stark contrast with negligible improvements in name space. The contrast echoes with the fact that the neural networks are much more effective in rank space compared to that in name space \autoref{fig: training_curve_NN}.}

\label{table: portfolio_performance_wo_tc}
\end{table}

\begin{table}[h!]
\scriptsize
\begin{tabular}{|c|lll|lll|lll|lll|}
\hline
\multirow{2}{*}{\textbf{year}} & \multicolumn{3}{c|}{\textbf{\begin{tabular}[c]{@{}c@{}}Name space\\ parametric model\end{tabular}}}       & \multicolumn{3}{c|}{\textbf{\begin{tabular}[c]{@{}c@{}}rank space \\ parametric model\end{tabular}}}      & \multicolumn{3}{c|}{\textbf{\begin{tabular}[c]{@{}c@{}}name space \\ neural networks\end{tabular}}}       & \multicolumn{3}{c|}{\textbf{\begin{tabular}[c]{@{}c@{}}rank space \\ neural networks\end{tabular}}}       \\ \cline{2-13} 
                               & \multicolumn{1}{c}{\textbf{return}} & \multicolumn{1}{c}{\textbf{vol}} & \multicolumn{1}{c|}{\textbf{SR}} & \multicolumn{1}{c}{\textbf{return}} & \multicolumn{1}{c}{\textbf{vol}} & \multicolumn{1}{c|}{\textbf{SR}} & \multicolumn{1}{c}{\textbf{return}} & \multicolumn{1}{c}{\textbf{vol}} & \multicolumn{1}{c|}{\textbf{SR}} & \multicolumn{1}{c}{\textbf{return}} & \multicolumn{1}{c}{\textbf{vol}} & \multicolumn{1}{c|}{\textbf{SR}} \\ \hline
\textbf{2007}                  & 1.63\%                              & 0.02                             & 0.79                             & -32.51\%                            & 0.04                             & -7.67                            & -15.51\%                            & 0.10                             & -1.61                            & 26.07\%                             & 0.10                             & 2.55                             \\
\textbf{2008}                  & 6.02\%                              & 0.04                             & 1.38                             & -43.78\%                            & 0.05                             & -9.58                            & -14.19\%                            & 0.13                             & -1.06                            & 36.40\%                             & 0.18                             & 2.04                             \\
\textbf{2009}                  & 5.71\%                              & 0.03                             & 1.65                             & -32.78\%                            & 0.04                             & -8.07                            & -3.13\%                             & 0.13                             & -0.23                            & 48.97\%                             & 0.13                             & 3.67                             \\
\textbf{2010}                  & -1.69\%                             & 0.02                             & -0.82                            & -32.83\%                            & 0.03                             & -12.68                           & -7.25\%                             & 0.08                             & -0.91                            & 43.14\%                             & 0.10                             & 4.32                             \\
\textbf{2011}                  & 0.71\%                              & 0.02                             & 0.31                             & -32.45\%                            & 0.03                             & -12.02                           & 5.10\%                              & 0.08                             & 0.66                             & 14.32\%                             & 0.10                             & 1.45                             \\
\textbf{2012}                  & -1.52\%                             & 0.02                             & -0.84                            & -26.53\%                            & 0.02                             & -11.84                           & -0.20\%                             & 0.07                             & -0.03                            & 20.41\%                             & 0.08                             & 2.42                             \\
\textbf{2013}                  & 0.02\%                              & 0.02                             & 0.01                             & -25.14\%                            & 0.02                             & -11.16                           & 0.21\%                              & 0.07                             & 0.03                             & 52.51\%                             & 0.10                             & 5.37                             \\
\textbf{2014}                  & 2.45\%                              & 0.02                             & 1.39                             & -22.52\%                            & 0.02                             & -9.74                            & -10.25\%                            & 0.07                             & -1.55                            & 35.55\%                             & 0.07                             & 4.76                             \\
\textbf{2015}                  & -0.31\%                             & 0.02                             & -0.16                            & -20.12\%                            & 0.03                             & -7.38                            & -6.89\%                             & 0.08                             & -0.89                            & 22.82\%                             & 0.10                             & 2.32                             \\
\textbf{2016}                  & 1.65\%                              & 0.02                             & 0.85                             & -17.68\%                            & 0.03                             & -5.45                            & 0.43\%                              & 0.10                             & 0.05                             & 56.09\%                             & 0.13                             & 4.31                             \\
\textbf{2017}                  & 0.36\%                              & 0.02                             & 0.22                             & -21.38\%                            & 0.02                             & -9.28                            & 2.62\%                              & 0.07                             & 0.36                             & 49.00\%                             & 0.09                             & 5.16                             \\
\textbf{2018}                  & 1.86\%                              & 0.02                             & 0.87                             & -28.83\%                            & 0.03                             & -11.45                           & -4.16\%                             & 0.06                             & -0.68                            & 27.94\%                             & 0.10                             & 2.81                             \\
\textbf{2019}                  & 3.34\%                              & 0.02                             & 1.81                             & -21.49\%                            & 0.02                             & -8.92                            & -13.52\%                            & 0.06                             & -2.12                            & 34.13\%                             & 0.10                             & 3.38                             \\
\textbf{2020}                  & 0.21\%                              & 0.04                             & 0.05                             & -42.12\%                            & 0.05                             & -9.02                            & -5.62\%                             & 0.09                             & -0.62                            & 56.62\%                             & 0.14                             & 4.14                             \\
\textbf{2021}                  & -2.06\%                             & 0.02                             & -0.86                            & -36.82\%                            & 0.03                             & -13.27                           & -9.69\%                             & 0.08                             & -1.27                            & 31.14\%                             & 0.13                             & 2.47                             \\
\textbf{2022}                  & -0.16\%                             & 0.03                             & -0.06                            & -33.01\%                            & 0.03                             & -11.71                           & 19.19\%                             & 0.09                             & 2.19                             & 15.69\%                             & 0.12                             & 1.26                             \\ \hline
\textbf{Avg}                   & 1.14\%                              & 0.02                             & 0.41                             & -29.37\%                            & 0.03                             & -9.95                            & -3.93\%                             & 0.08                             & -0.48                            & 35.68\%                             & 0.11                             & 3.28                             \\ \hline
\end{tabular}
\caption{\textbf{Portfolio performance with transaction costs.} The portfolio performances in rank space degrade dramatically, for both the parametric model and the neural networks. The substantial degradation arises from the substantial costs associated with realizing rank return in continuous time limit through intraday rebalancing. Nevertheless, the portfolio calculated by neural networks in rank space still yields good results, as the significant transaction costs are compensated by the impressive returns and Sharpe ratio in \autoref{table: portfolio_performance_wo_tc}, column 4. We choose 2 basis points to account for the transaction costs from bid-ask spread.}
\label{table: portfolio_performance_with_tc}
\end{table}

\subsection{Dollar neutrality}\label{sec: portfolio_performance_market_dollar_neutrality}
\begin{figure}[h!]
    \centering
    \includegraphics[width=\columnwidth]{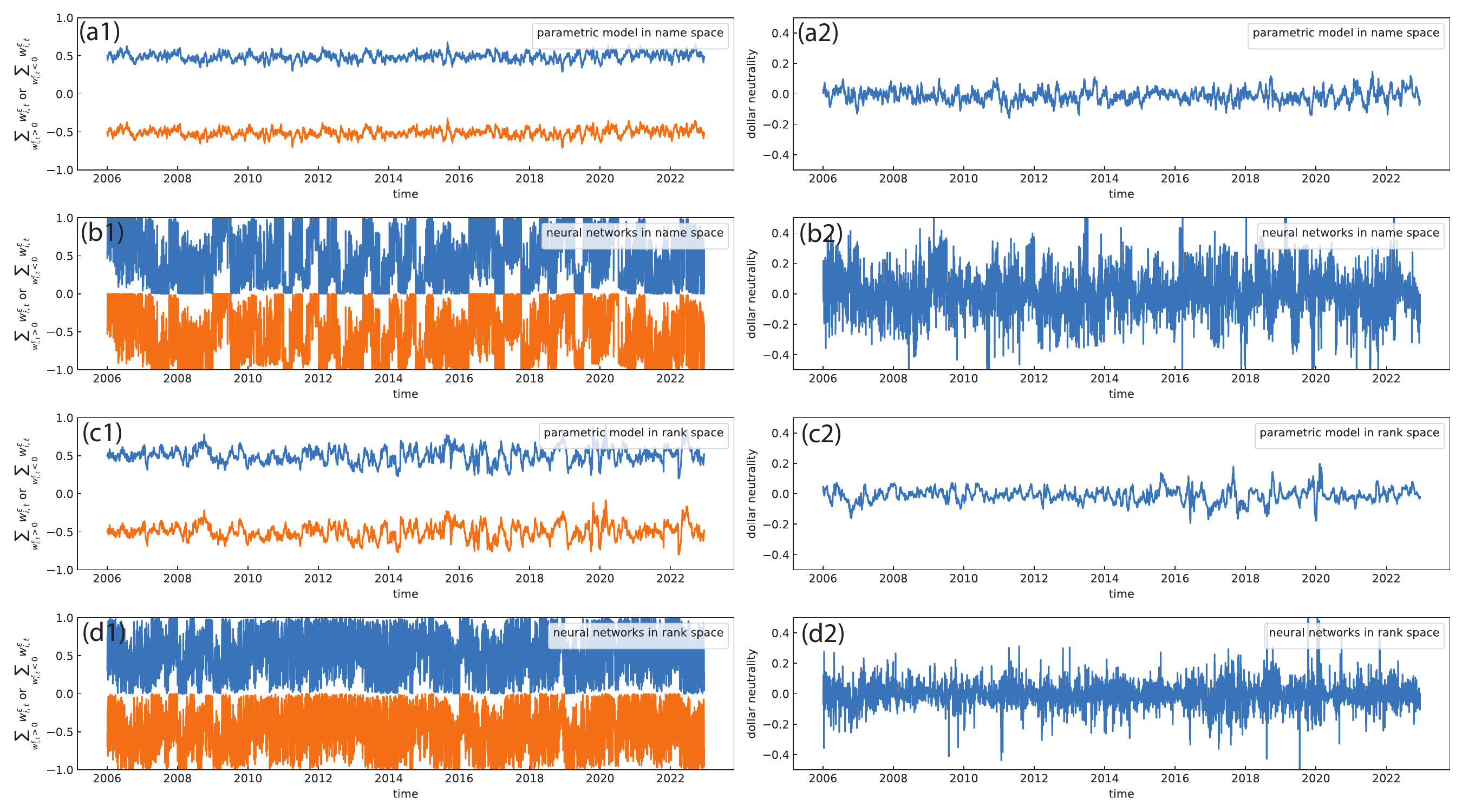}
    \caption{\textbf{Portfolio weights in residual return and dollar neutrality. (a1-d1)} The temporal dependence of average long/short portfolio weights in residual space, $\sum_{w_{i,t}^\epsilon>0}w_{i,t}^\epsilon$ and $\sum_{w_{i,t}^\epsilon<0}w_{i,t}^\epsilon$. \textbf{(a2-d2)} The deviation from dollar neutrality measured by $\frac{\sum_i w_{i,t}^R}{\sum_i |w_{i,t}^R|}$. We consider four scenarios: (a1-a2) parametric model in name space; (b1-b2) neural networks in name space; (c1-c2) parametric model in rank space; (d1-d2) neural networks in rank space.}
    \label{fig: dollar_neutrality}
\end{figure}

We characterize the long or short proportion of the portfolio weights in equity space, $\sum_{i: w_{i,t}^R>0}w_{i,t}$ or $\sum_{i: w_{i,t}^R<0}w_{i,t}$ and the dollar neutrality, $\frac{\sum_{i}w_{i,t}^R}{\sum_{i}|w_{i,t}^R|}$. The results are presented in \autoref{fig: dollar_neutrality}(a1-d1) and \autoref{fig: dollar_neutrality}(a2-d2), where we consider $w_t^R$ calculated by four scenarios: (i) the parametric benchmark model in name space (\autoref{fig: dollar_neutrality}(a1, a2), (ii) DNNs in name space (\autoref{fig: dollar_neutrality}(b1, b2), (iii) the parametric benchmark model in rank space (\autoref{fig: dollar_neutrality}(c1, c2), (iv) DNNs in rank spce (\autoref{fig: dollar_neutrality}(d1, d2)). Notably, the long or short proportion of $w_t^R$ by neural networks (\autoref{fig: dollar_neutrality}(b1, d1)) is much more volatile than those by the parametric model (\autoref{fig: dollar_neutrality}(a1, c1)), as a result of flexible leverage adopted by neural networks. However, the dollar neutrality is satisfied on average thanks to the market neutrality of the portfolios.

\subsection{Dependence on transaction costs}\label{sec: portfolio_performance_transaction_costs}

\begin{figure}[h!]
    \centering
    \includegraphics[width=\columnwidth]{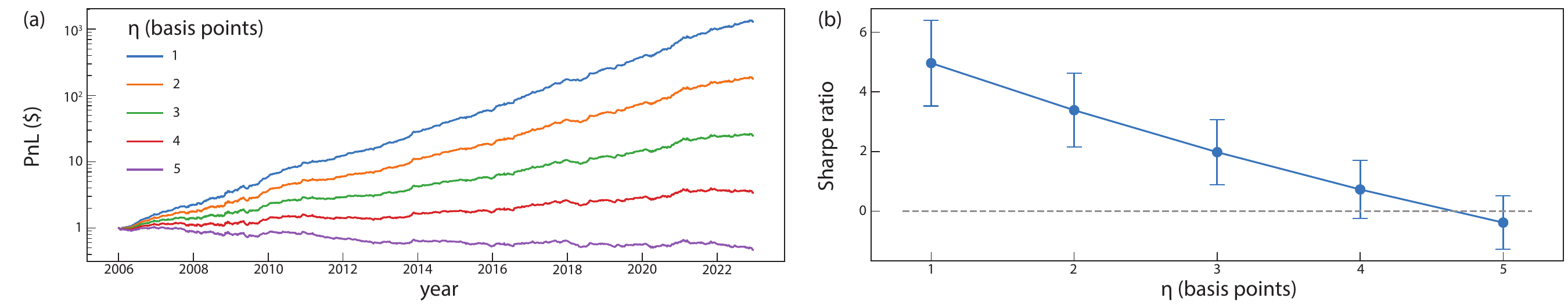}
    \caption{\textbf{PnL dependence on transaction cost factor $\eta$. (a)} The PnL with varing levels of the transaction cost factor $\eta$. The underlying portfolio weights are derived from the neural networks in rank space. \textbf{(b)} The average Sharpe ratio from 2006 to 2022 at different values of $\eta$ derived from (a). The strategy stops to profit with 5 basis points transaction costs due to substantial costs associated with realizing rank returns in continuous time limit. The significant change in Sharpe ratio under varying transaction costs underscores the strategy's sensitivity to transaction costs and motivates ongoing improvements in "trading ranks".}
    \label{fig: transaction_cost_dependence}
\end{figure}

We present the sensitivity of portfolio performance to transaction costs. We show the PnL with different transaction cost factor $\eta$ in \autoref{fig: transaction_cost_dependence}(a) and the corresponding Sharpe ratio in \autoref{fig: transaction_cost_dependence}. The current strategy shows significant sensitivity to the transaction cost factor and stops to profit with $\eta=5$ basis points. A more effective strategy to realize return in rank space $\Tilde{r}_t$ will help the strategy more immune to transaction costs.

\subsection{A characteristic time between rank switching}\label{sec: characteristic_time_rank_swap}
\begin{figure}[h!]
    \centering
    \includegraphics[width=\columnwidth]{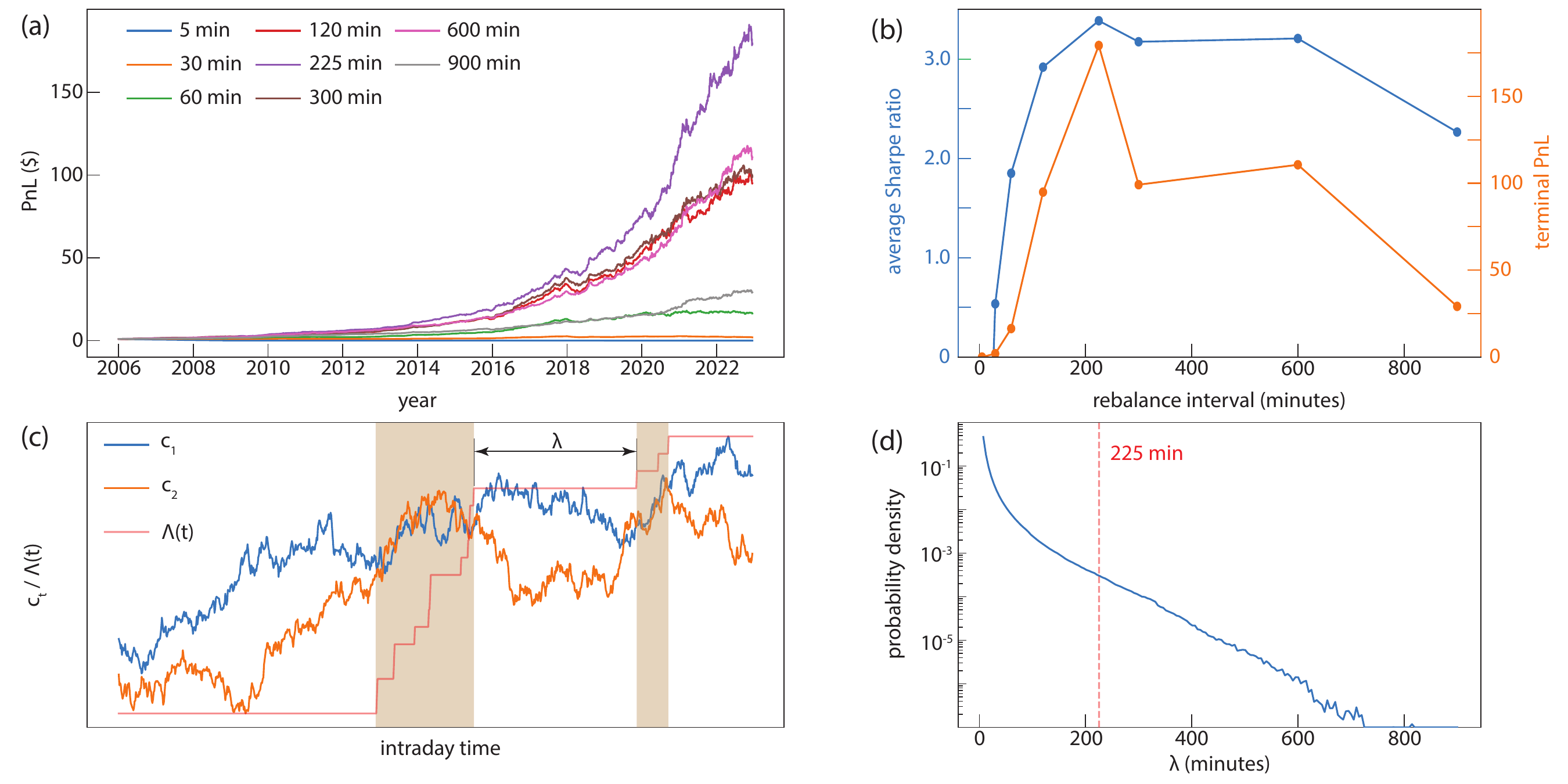}
    \caption{\textbf{PnL dependence on rebalancing interval and characteristic rank-swapping time. (a)} The PnL across various intraday rebalancing intervals $\mathcal{T}$ to realize the rank return in continuous time limit $\Tilde{r}_t$. The underlying portfolio weights are derived from the neural networks in rank space. \textbf{(b)} Averaged Sharpe ratio between 2006 and 2022 and the terminal PnL as functions of rebalance intervals from (a). Both metrics peak at $\mathcal{T}=225$ minutes. \textbf{(c)} Schematic representation of two capitalization processes,$c_1$ and $c_2$. The $\Lambda(t)$ measures the cumulative time that the capitalization processes $c_1$ and $c_2$ cross. Rank switching time $\lambda$ is the interval between the increments in local time. This stochastic system has two characteristic regimes: (i) the idle regime, where the two capitalization processes are distant, maintaining constant $\Lambda(t)$ with prolonged $\lambda$; (ii) the collision regime (highlighted in brown shaded area in \autoref{fig: rebalance_interval_dependence}(c)), where the two capitalization processes stay close, leading to rapid increases in $\Lambda(t)$ and short $\lambda$. \textbf{(e)} The empirical distributions of rank-swapping time $\tau$ based on the intraday market data. Low (high) $\lambda$ arises from the "idle" ("collision") regime. The red dashed line marks the optimal rebalancing interval, $\mathcal{T}=225$ minutes, positioned at the intersection between the idle and collision regimes, suggesting it is a balanced choice for minimizing transaction costs while responding effectively with rank-swapping events.}
    \label{fig: rebalance_interval_dependence}
\end{figure}

The rebalancing interval, $\mathcal{T}$, turns out to be a crucial parameter for statistical arbitrage portfolios in rank space. To demonstrate, we present the calculated PnL with varying rebalancing intervals in \autoref{fig: rebalance_interval_dependence}(a), where the portfolio weights are calculated from the neural networks in rank space. The associated average Sharpe ratio and terminal PnL as a function of the rebalancing interval are summarized in \autoref{fig: rebalance_interval_dependence}(b), with both peaking at 225 minutes.

Here, we delve into the rationale behind the optimal 225-minute interval, starting with an examination of two proximate capitalization processes modeled as Brownian motions, $c_{1,t}$ and $c_{2,t}$ (\autoref{fig: rebalance_interval_dependence}(c)). Given that transaction costs from intraday rebalancing primarily arise from rank swaps in capitalization (section 2.3), we measure the cumulative time $\Lambda(t)$ that the capitalization processes cross (\autoref{fig: rebalance_interval_dependence}(c), red line),
\begin{equation}
\Lambda(t) = \lim_{\delta\downarrow 0}\int_0^t \boldsymbol{1}_{\{|c_{1,\tau} - c_{2,\tau}|\leq \delta\}}d\tau
\end{equation}
The rank-swapping interval $\lambda$ is defined as the time between the cross of capitalization processes, or equivalently, the increments of $\Lambda(t)$. This classical interacting Brownian system features two characteristic regimes: (i) the idle regime, where the two capitalization processes are distant, maintaining constant $\Lambda(t)$ with prolonged $\lambda$; (ii) the collision regime (highlighted in brown shaded area in \autoref{fig: rebalance_interval_dependence}(c)), where the two capitalization processes stay close, leading to rapid increases in $\Lambda(t)$ and short $\lambda$. We show the empirical distribution of $\lambda$ on real market in \autoref{fig: rebalance_interval_dependence}(d), where the small $\lambda$ values arise from the collision regime and larger $\lambda$ values from the idle regime, following approximately an exponential distribution as a typical signature for standard Brownian particle systems. The 225 minutes is situated at the intersection of the two regimes, establishing it as a characteristic time for rank switching.

In the detailed analysis of the intraday rebalancing in \eqref{eq: cost_intraday_rebalance} and \autoref{fig: schematic_intraday_dependence_frequency_dependence}, the transaction costs arise from (i) latency costs due to delayed reactions post-rank-swapping, and (ii) costs from bid-ask spreads incurred during active trading. For the collision regime in \autoref{fig: rebalance_interval_dependence}(c), it is preferable to delay trading to minimize bid-ask spread costs. Conversely, in the idle regime, immediate trading is preferable to reduce latency costs. The 225-minute interval effectively differentiates these regimes, thus optimizing overall transaction costs.

The discussion above highlights the challenge in trading ranks -- discerning between the collision and idle regimes and trading at their intersection. Our current intraday rebalancing approach crudely harnesses average behavior of U.S. equity market, and leaves considerable scope for enhancement that we will follow up on in subsequent research papers.

\clearpage
\bibliographystyle{plainnat}
\bibliography{reference}

\end{document}